\documentclass{article}

\usepackage{arxiv}

\usepackage[utf8]{inputenc} % allow utf-8 input
\usepackage[T1]{fontenc}    % use 8-bit T1 fonts
\usepackage[hidelinks]{hyperref}       % hyperlinks
\usepackage{url}            % simple URL typesetting
\usepackage{booktabs}       % professional-quality tables
\usepackage{amsfonts}       % blackboard math symbols
\usepackage{nicefrac}       % compact symbols for 1/2, etc.
\usepackage{microtype}      % microtypography
\usepackage{lipsum}
\usepackage{natbib}
\usepackage{caption} %set the size of the figure caption
\usepackage{amssymb} %math symbol
\usepackage{amsmath} %math
\usepackage{mathrsfs} % Raph Smith?s For­mal Script 
\usepackage{graphicx}
\usepackage{enumerate}
\usepackage{url} % not crucial - just used below for the URL 
\usepackage[amsmath,amsthm,thmmarks]{ntheorem}
\usepackage{bbm} %indicator function

\usepackage{subcaption} %align figures
\usepackage{appendix}
\usepackage{algorithm} %algorithm environment
\usepackage{algpseudocode}
\usepackage{booktabs} %set the head line and below line of the table

\newtheorem{theorem}{Theorem}
\newtheorem{lemma}{Lemma}
\newtheorem{assumption}{Assumption}
\newtheorem{remark}{Remark}

\newtheorem{definition}{Definition}

\DeclareMathOperator*{\argmax}{arg\,max}
\DeclareMathOperator*{\argmin}{arg\,min}

\title{Generalized Geographically Weighted Regression Model within a Modularized Bayesian Framework}

\author{
  Yang Liu\thanks{Correspondence: 
    Yang Liu, MRC Biostatistics Unit, University of Cambridge, Robinson Way, CB2 0SR} \\
  MRC Biostatistics Unit\\
  University of Cambridge\\
  Cambridge, UK \\
  \texttt{yang.liu@mrc-bsu.cam.ac.uk} \\
   \And
 Robert J.B. Goudie \\
  MRC Biostatistics Unit\\
  University of Cambridge\\
  Cambridge, UK \\
  \texttt{robert.goudie@mrc-bsu.cam.ac.uk} \\
}

\begin{document}
\maketitle

\begin{abstract}
Geographically weighted regression (GWR) models handle geographical dependence through a spatially varying coefficient model and have been widely used in applied science, but its general Bayesian extension is unclear because it involves a weighted log-likelihood which does not imply a probability distribution on data. We present a Bayesian GWR model and show that its essence is dealing with partial misspecification of the model. Current modularized Bayesian inference models accommodate partial misspecification from a single component of the model. We extend these models to handle partial misspecification in more than one component of the model, as required for our Bayesian GWR model. Information from the various spatial locations is manipulated via a geographically weighted kernel and the optimal manipulation is chosen according to a Kullback–Leibler (KL) divergence. We justify the model via an information risk minimization approach and show the consistency of the proposed estimator in terms of a geographically weighted KL divergence.
\end{abstract}

% keywords can be removed

\section{Introduction}
Conventional regression models have been widely used in various studies to infer the association between variables. While basic regression models often assume an independent sampling scheme, geographical dependence must be taken into consideration when the dataset or sampling scheme has a spatial structure. Therefore, rather than assuming a constant association between variables with constant coefficients, models with geographically-variable coefficients have been proposed for this purpose. Suppose we have observations $(X_i,Y_i)$ at sampling location $i$ with coordinates $(u_i,v_i)$, $i=0,\cdots,n$. We assume that the unknown true data generating process of the outcome $Y_i$, given the covariate vector $X_i$, is $\check{p}_i(Y_i|X_i)$ at a particular location $i$. To model $\check{p}_i$, we assume a generalized linear model (GLM) $\mathbb{E}(Y_i|X_i)= g^{-1}(X_i\varphi(u_i,v_i))$, with link function $g$, and where the coefficient $\varphi(u,v)$ is a smooth function with respect to $(u,v)$. For simplicity, we define $\varphi_i\equiv\varphi(u_i,v_i)$ for location $i$.

In addition to the coefficient $\varphi_i$, for some generalized linear regression models, such as negative binomial or beta regression, for each location $i$ there is an additional parameter $\theta_i$ that determines the variability (scale) of the distribution. The additional parameter $\theta_i$ is usually regarded as a nuisance parameter. This variability could be attributed to sampling or measurement errors, which may be different at different locations. We assume that $\theta_i$ is similar, but not the same, across spatial locations but the variability is not spatially smooth. For instance, consider a variability induced by a difference in measurement equipment: each location may have arbitrarily used different measurement equipment, and consequently the variabilities of observations at different locations are not constant but also not spatially smooth. We denote the likelihood of the GLM as $p(Y_i|\theta_i,\varphi_i)$ at location $i$. For example, 
% abandoned
\if 0 in the case of a Gaussian distribution when an identity link function is used, the likelihood is:
\begin{equation}
p(Y_i|\theta_i,\varphi_i) = \frac{1}{\theta_i\sqrt{2\pi}}\exp\left(-\frac{1}{2{\theta_i}^2}(Y_i-X_i\varphi_i)^2\right),
\end{equation}
and \fi in the case of a negative binomial likelihood, with a log link function, the likelihood is:
\begin{equation}
\resizebox{0.9\textwidth}{!}{$p(Y_i|\theta_i,\varphi_i) = \frac{\Gamma(Y_i+{\theta_i}^{-1})}{\Gamma({\theta_i}^{-1})\Gamma(Y_i+1)}\left(\frac{1}{1+\theta_i\exp(X_i\varphi_i)}\right)^{{\theta_i}^{-1}} \left(\frac{\theta_i\exp(X_i\varphi_i)}{1+\theta_i\exp(X_i\varphi_i)}\right)^{Y_i}.$}
\label{E2}
\end{equation}
We assume that a single location $i=0$ is of primary interest, and our first aim is to estimate $\varphi_0$ and $\theta_0$ at this location. We will then consider the case when multiple locations are of interest.

Several modelling approaches have been proposed for geographically variable coefficients. One class of approaches involves clustering locations into groups and considering a group-wise estimation of the coefficients. For example, \cite{doi:10.1080/01621459.2018.1529595} proposed spatially clustered coefficient (SCC) regression that adds a penalty term to the residual sum of squares such that differences of coefficients for neighbouring locations are penalized and consequently locations may share the same coefficient. \cite{SUGASAWA2021100525} proposed a partially clustered regression that allocates locations into groups, with locations sharing the same coefficients within a group.
Another class of approaches are the Bayesian spatially varying coefficient (SVC) model \citep{doi:10.1198/016214503000170} and its extensions \citep[e.g.,][]{paez2005interpolation, finley2007spbayes, berrocal2010spatio, https://doi.org/10.1111/j.1541-0420.2009.01333.x}. These have been developed within a standard Bayesian framework with geographically varying associations $\mathbb{E}(Y_i|X_i)= X_i\varphi_i$. SVC models induce geographical dependence via a random spatial adjustment to coefficients, such as $\varphi_i = \varphi_{fix} + \varphi_{random,i}$, where $\{\varphi_{random,i}\}_{i=1}^n$ are modeled by a Gaussian random field with a covariance structure corresponding to the geographical dependence of $n$ locations. SVC models share the power of hierarchical modeling \citep{doi:10.1146/annurev-statistics-060116-054155} via their similarity to spatial hierarchical models, which use a random Gaussian field to model the regression error \citep[e.g.,][]{zhu2005modeling,lin2010estimating,doi:10.1080/02664763.2011.570315,10.2307/24311007,doi:10.1177/0962280218797362,MARQUES2020100386}. SVC models do not involve any geographical weight function so the probability density is always proper and standard Bayesian inference can be applied. However, the sampling of the posterior distribution under SVC models can be challenging because the dimension of parameters (i.e., $\{\varphi_{random,i}\}_{i=1}^n$) increases with the number of locations, making both sampling parameters and inverting the spatial covariance matrix computationally difficult. This issue can be avoided in the Gaussian case because $\{\varphi_{random,i}\}_{i=1}^n$ can be integrated out to obtain the marginal likelihood of $\varphi_{fix}$ explicitly and the conditional posterior of $\{\varphi_{random,i}\}_{i=1}^n$ given $\varphi_{fix}$ is analytically tractable. However, in generalized linear models marginalization of $\{\varphi_{random,i}\}_{i=1}^n$ is not usually feasible, e.g. for binary and Poisson \citep{https://doi.org/10.1111/j.1467-9868.2008.00663.x}, and so sampling and computation could be problematic in practice due to the high dimension of the parameters if we have lots of locations \citep{SUGASAWA2021100525}. Therefore, literature about the SVC model for generalized linear models is sparse and the model may not be computationally practicable.

Other attractive and simpler alternatives are geographically weighted regression (GWR) models \citep[e.g.,][]{doi:10.1080/02693799608902100,doi:10.1111/j.1538-4632.1996.tb00936.x} and its extensions \citep[e.g.,][]{doi:10.1002/sim.2129,https://doi.org/10.1111/j.1538-4632.2012.00841.x, da2014geographically, DASILVA2017279, https://doi.org/10.1002/env.2485, liu2018geographically,doi:10.1080/13658816.2020.1720692, TASYUREK2020258, doi:10.1080/13658816.2020.1775836}, which have been widely adopted in many spatial application areas \citep[e.g.,][]{10.1093/icesjms/fsp224,7517288,MAYFIELD2018e223,WANG201995,WU2020121089,Mohammed33}. GWR models use the first law of geography to justify additionally using data that are sampled from neighbouring locations when we have insufficient samples at a location of interest $i=0$ to accurately estimate parameters at this location using only data from this location. The first law of geography states that `everything is related to everything else, but near things are more related than distant things' \citep[][]{tobler1970computer}. ``Borrowing'' samples from neighbouring locations to support the estimation of $\varphi_0$ should decrease the variance of estimates, although bias might be introduced.

For now, assume we have $m$ observations $Y_{i,1:m}=(Y_{i,1},Y_{i,2},\cdots,Y_{i,m})$ at each location $i$, with $i=0,\cdots,n$. The complete set of observations is $Y_{0:n,1:m} = (Y_{0,1:m}, Y_{1,1:m}, \allowbreak\dots,\allowbreak Y_{n,1:m})$ with corresponding location-specific parameters $\theta_{0:n}=(\theta_0,\cdots,\theta_n)$. We assume that $Y_{i,1},\cdots,Y_{i,m}$ are independent identical observations of the random variable $y_i$ at location $i$. In addition, we assume $Y_{0,1:m}, \allowbreak\dots,\allowbreak Y_{n,1:m}$ are independent but not necessarily identically-distributed. Let $d_i$ be the geographic distance between location of interest $0$ and location $i$. The generalized GWR likelihood is a locally-weighted likelihood:
\begin{equation}
p(Y_{0:n,1:m}|\theta_{0:n},\varphi_0)=p(Y_{0,1:m}|\theta_0,\varphi_0)  \prod_{i=1}^n  p(Y_{i,1:m}|\theta_i,\varphi_0 )^{W(d_i,\eta)},
\label{E3}
\end{equation}
with coefficient $\varphi_0$ and where $W(d_i,\eta)$ is a geographically weighted kernel, with bandwidth $\eta$, determined by the distance. Following the first law of geography, geographically weighted kernels gradually decrease to 0 as the distance $d_i$ increases. One popular choice of weighted kernel is a Gaussian kernel \citep{doi:10.1111/j.1538-4632.1996.tb00936.x}
\begin{equation}
W(d_i,\eta) = \exp\left(-\frac{d_i^2}{\eta^2}\right),
\label{E4}
\end{equation}
where $\eta$ is a geographical bandwidth which regulates the kernel size.

Inference for GWR models has usually been conducted in a frequentist framework, but a Bayesian extension of the GWR model would allow introduction of prior information, and also simplify situations where the covariance of the estimator is not easily obtainable. However, Bayesian inference for general GWR models is not immediately clear since, \eqref{E3} is not in general a proper probability density if the power terms are not 1. Hence, Bayes' theorem does not apply. In the special case of a Gaussian likelihood, $W(d_i,\eta)$ can be viewed as a scale parameter of $\theta_i$ and thus we obtain a proper probability density. This special case has previously been considered, allowing inference for Gaussian GWR models within a standard Bayesian framework \citep{SUBEDI20182574011,doi:10.1177/0160017620959823}. However, a Bayesian extension for a broader distribution family is unclear, and to the best of our knowledge, no previous papers have considered this problem.

In this article, we extend the generalized GWR model to the Bayesian framework and justify its usage. Observe that \eqref{E3}, ignoring the power terms, treats data sampled from neighbouring locations $i$, $i\neq0$, as if they share the same relationship with covariate $X_i$ as data sampled from the location of interest $i=0$. This inevitably leads to the problem of misspecification since $\varphi_i \neq \varphi_0$ due to the spatial non-stationarity. The degree of misspecification depends on the total variation of $\varphi_i$. This observation suggests that the essence of the Bayesian GWR model is dealing with misspecification due to incorporating extra observations from neighbouring locations and inspired us to draw ideas from the literature considering partial misspecification of Bayesian models and the modularized Bayesian analysis \citep{liu2009modularization}. The model involves a geographically powered posterior, with the power term being a deterministic functional form of the geographical distance. The contribution from each location to the inference of the parameter of interest is manipulated through a geographical bandwidth in the power term and we discuss the optimal selection of this bandwidth so that the negative impact from misspecification and positive impact from extra observations are well balanced. We show some theoretical properties of the model and outline the algorithm.

\section{Robust Bayesian Inference and Modularization}
% abandoned
\if 0 Consider the observed data $Y$ and we assume a likelihood subject to parameters of interest $\psi\in\Psi$ as $p_\psi(Y|\psi)$. The inference of parameters of interest $\psi$ is obtained by integrating information from both prior knowledge $\pi(\psi)$ and data $Y$ through Bayes theorem $p(\psi|Y)\propto p_\psi(Y|\psi)\pi(\psi)$. The justification of the Bayesian inference relies on the correct specification of the model (in particular, the likelihood of data $p_\psi(Y|\psi)$ with respect to the parameters of interest). Here, we adopt the M-closed view that a model is correctly specified if the true data generating process $\check{p}(Y)$ is exactly equal to a parametric distribution $p_{\psi_0}(Y|\psi_0)$, $\psi_0\in\Psi$, which is subsequently referred as the likelihood \citep{https://doi.org/10.1111/rssb.12158}. In contrast, given a likelihood with respect to parameters of interest in a form of parametric distributions, the M-open view states that there is no such parametric distributions subject to parameters of interest where data is generated (i.e., $\{\psi: p_\psi(Y|\psi) = \check{p}(Y)\}=\emptyset$).\fi

Several attractive properties of Bayesian inference rely on the correct specification of the model. However, it is generally impossible to ensure the correct specification of a complete Bayesian model. Here, we adopt the M-closed view that a model is correctly specified if the true data generating process $\check{p}(Y)$ is exactly equal to a parametric distribution $p_{\psi_0}(Y|\psi_0)$, given parameters $\psi_0\in\Psi$, which is subsequently referred as the likelihood \citep{https://doi.org/10.1111/rssb.12158}. Misspecification might exist in all aspects of the model, or in only a few components \citep[or modules in the terminology of][]{liu2009modularization} of the model. 

In the case of all aspects of the model being misspecified, modification of the conventional Bayesian model is required to improve the robustness of the model. One approach is to raise the likelihood to a power term and regard its logarithm as a loss function \citep{https://doi.org/10.1111/j.1467-9868.2007.00650.x,https://doi.org/10.1111/rssb.12158,10.1093/biomet/asx010}, to obtain a weighted likelihood similar to the generalized GWR in \eqref{E3}:
\begin{equation}
p_{\text{pow},\eta}(\psi|Y) \propto p_\psi(Y|\psi)^\eta \pi(\psi).
\label{E9}
\end{equation} 
This is called the power posterior or fractional posterior, with power $\eta$. While weighted likelihoods have a long history in frequentist statistics \citep[e.g.,][]{doi:10.1080/01621459.2000.10474280,https://doi.org/10.1111/j.0006-341X.2000.00483.x,https://doi.org/10.2307/3316141,https://doi.org/10.1002/sta4.80}, it is only recently that justification of their usage in Bayesian statistics has been studied. One interpretation of the power term is that it adjusts the sample size with a multiplier $\eta$ \citep{doi:10.1080/01621459.2018.1469995}. Another interpretation is that it is equivalent to a data-dependent prior \citep{martin2017}. \cite{doi:10.1080/01621459.2018.1469995} further argue that \eqref{E9} approximates $p(\psi|\mathbb{D}_{KL}(p_\psi(\cdot|\psi), \check{p}(\cdot))<R)$ under mild conditions, where the Kullback-Leibler (KL) divergence $\mathbb{D}_{KL}(p_\psi(\cdot|\psi), \check{p}(\cdot))=\int \check{p}(y) \log(\check{p}(y)/p_\psi(y|\psi))dy$ and $R$ is determined by the number of samples and the power $\eta$. The contraction of the power posterior is shown by \cite{bhattacharya2019}. These papers suggest that, in the case of a M-open view, where the true data generating process does not belong to the parametric distributions termed as likelihood, inference can proceed by looking for parameters whose likelihood approximates the true data generating process. In addition, an appropriate choice of $\eta$ can accommodate this departure of misspecified $p_\psi(Y|\psi)$ from the truth $\check{p}(Y)$ and the model is robust \citep{doi:10.1080/01621459.2018.1469995}. Importantly, the power $\eta$ controls the relative credence given to the observed data and the prior; consequently it is not deemed as a parameter. Therefore, a prior is not assigned for $\eta$ and it is not updated via Bayes theorem. 

\begin{figure}[t] 
\setlength{\abovecaptionskip}{0cm}
\setlength{\belowcaptionskip}{0cm}
\centering 
\includegraphics[scale=0.2]{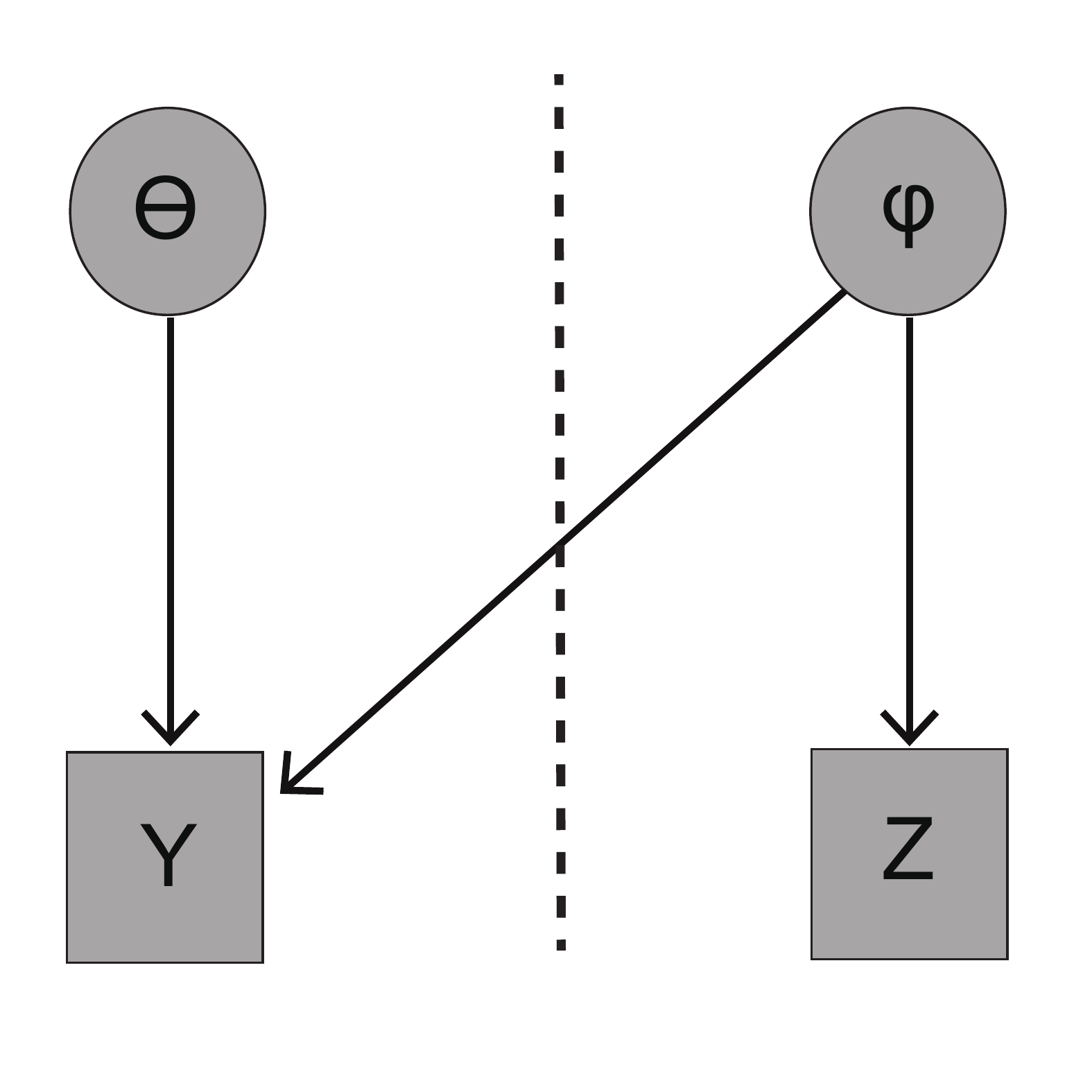} 
\caption[DAG representation of a two module model.]{\textbf{DAG representation of a two module model.} The modules are separated by a dashed line.} 
\label{F1}
\end{figure}

In the case of partial misspecification, misspecification of even a single module can cause incorrect estimation of other modules, even if these modules are correctly specified \citep{plummer2015cuts, liu2020stochastic}. Consider the two module model illustrated in Figure \ref{F1}, with likelihood terms $p(Y|\theta,\varphi)$ and $p(Z|\varphi)$, and prior terms $\pi(\theta)$ and $\pi(\varphi)$. The posterior distribution, with parameters of interest $\psi=(\theta,\varphi)$, is
\[
    p(\psi|Y,Z)=p(\theta|Y,\varphi)p(\varphi|Y,Z) = \frac{p(Y|\theta,\varphi)\pi(\theta)}{p(Y|\varphi)}\frac{p(Y|\varphi)p(Z|\varphi)\pi(\varphi)}{p(Y,Z)}.
\]
Suppose that the specification of the likelihood for $Y$ is suspected to be incorrect. If we wish to prevent $Y$ affecting estimation of $\varphi$, then we can use the cut distribution \citep{lunn2009combining}, defined for this model as
\[
    p_{\text{cut}}(\psi|Y,Z) := p(\theta|Y,\varphi)p(\varphi|Z) = \frac{p(Y|\theta,\varphi)\pi(\theta)}{p(Y|\varphi)}\frac{p(Z|\varphi)\pi(\varphi)}{p(Z)},
\]
Note that under the cut distribution $\varphi$ depends on only the data $Z$; the data $Y$ makes no contribution to the estimation of $\varphi$. This is called ``cutting the feedback'' \citep{lunn2009combining}. This model has been used for Bayesian propensity scores \citep[e.g.,][]{McCandlessDouglasEvansSmeeth2010,kaplan2012two,doi:10.1080/01621459.2013.869498} where feedback from the outcome module to the propensity score module should be removed \citep{rubin2008,zigler2013model}. It has also been used in various other fields \citep[e.g.,][]{BLANGIARDO2011379,arendt2012quantification,https://doi.org/10.1029/2018WR023054}.

The cut distribution and the standard posterior are two extremes: all information from the suspect module is either removed or retained. However, completely cutting or retaining the feedback from the suspect module might either lose usable information or introduce excessive bias. To control the feedback from the potentially misspecified module, a combination of the power posterior and cut model was recently proposed by \cite{carmona2020semi}. Their Semi-Modular Inference (SMI) model introduces an auxiliary variable $\tilde{\theta}$, which has the same distribution as $\theta$, to regulate the contributions to the estimation of $\varphi$. Given a prior $\pi(\varphi,\tilde{\theta})$, the SMI distribution of the augmented parameter $(\theta,\tilde{\theta},\varphi)$ is
\[
    p_{\eta}(\theta,\tilde{\theta},\varphi|Y,Z)=p_{\text{pow},\eta}(\tilde{\theta},\varphi|Y,Z)p(\theta|Y,\varphi),
\]
where
\[
    p_{\text{pow},\eta}(\tilde{\theta},\varphi|Y,Z)\propto p(Z|\varphi)p(Y|\varphi,\tilde{\theta})^\eta \pi(\varphi,\tilde{\theta})
\]
is a power posterior of $\tilde{\theta}$ and $\varphi$, with power $\eta$. The SMI distribution of the parameters of interest $\psi=(\theta,\varphi)$ is
\[
    p_{\eta}(\psi|Y,Z)=\int p_{\eta}(\theta,\tilde{\theta},\varphi|Y,Z) d\tilde{\theta}.
\]
The power $\eta$ controls how much information from the suspect module involving $Y$ is used to estimate $\varphi$.

\section{Modularized Bayesian Inference for Multiple Modules}
\label{SEC3}
\subsection{Standard Bayesian posterior and cut distribution}
\begin{figure}[t] 
\setlength{\abovecaptionskip}{0cm}
\setlength{\belowcaptionskip}{0cm}
\centering 
\includegraphics[scale=0.5]{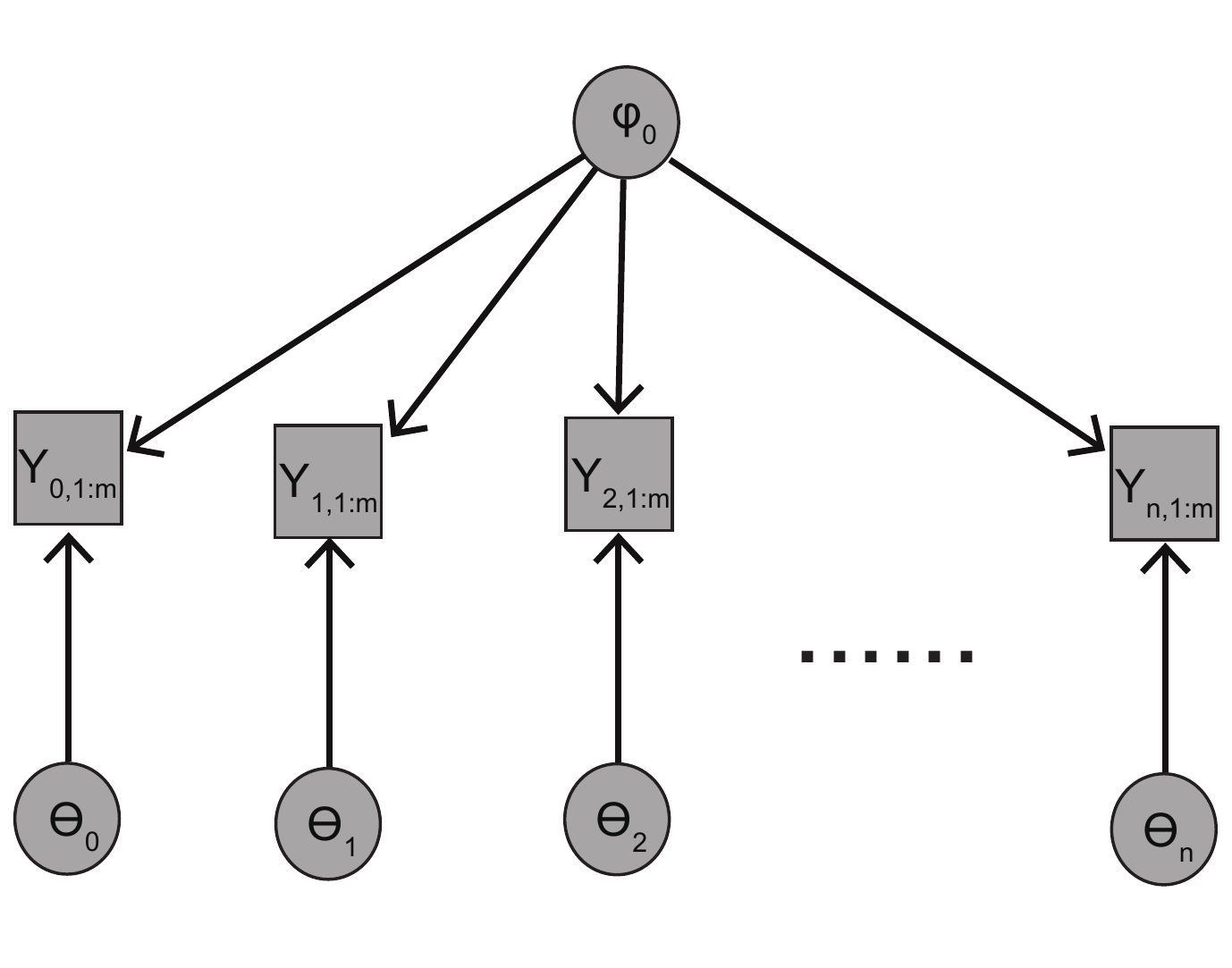} 
\caption[DAG representation when $\varphi(u_i,v_i)=\varphi_0$.]{\textbf{DAG representation when $\varphi(u_i,v_i)=\varphi_0$.}} 
\label{F2}
\end{figure}

To establish notation, first consider the simple case when the spatial coefficient function $\varphi(u_i,v_i)=\varphi(u_0,v_0)=\varphi_0$; that is $\varphi$ is constant across the whole geographical space and so we can directly include all data from all locations into the model. Denote the likelihood $p(Y_{i,1:m}|\theta_i,\varphi_0)$ at location $i$, with $i=0,1,\cdots,n$. The DAG of this model is shown in Figure \ref{F2}. The joint distribution with an independent prior $\pi(\theta_{0:n},\varphi_0)=\pi(\varphi_0) \prod_{i=0}^n \pi(\theta_i)$ is
\[
p(Y_{0:n,1:m}, \theta_{0:n},\varphi_0)= \pi(\theta_{0:n},\varphi_0) \prod_{i=0}^n p(Y_{i,1:m}|\theta_i,\varphi_0).
\]

The following lemma gives the form of the standard Bayesian posterior.
\begin{lemma}
The standard Bayesian posterior is:
\begin{equation}
p(\theta_{0:n},\varphi_0|Y_{0:n,1:m})=p(\theta_0,\varphi_0|Y_{0:n,1:m}) \prod_{i=1}^n p(\theta_i|Y_{i,1:m},\varphi_0).
\label{E10}
\end{equation}
\end{lemma}
\begin{proof}
See appendix.
\end{proof}
Note that, estimation of $(\theta_0,\varphi_0)$ is influenced by all observations $Y_{0,1:m}, \dots,\allowbreak Y_{n,1:m}$ as is standard in Bayesian inference: the contribution from any location is equal in the sense that no manipulation of feedback is conducted.

In contrast, consider the case when $\varphi(u_i,v_i)$ is not constant. If we nevertheless include data from location $(u_i,v_i)$, $i\neq 0$ to estimate the parameter $(\theta_0,\varphi_0)$ and regard $y_i\sim p(\cdot|\theta_i,\varphi_0)$ as module $i$, $i = 0,1,\cdots,n$, then the likelihood $\Pi_{i=0}^{n}p(Y_{i,1:m}|\theta_i,\varphi_0)$ is clearly misspecified since $\varphi_0\neq \varphi(u_i,v_i)$. A straightforward way to handle this misspecification is to remove the influence of these modules on the estimation of $\varphi_0$ by using the cut distribution. The cut distribution for this model is:
\begin{equation}
\begin{aligned}
p_{\text{cut}}(\theta_{0:n},\varphi_0|Y_{0:n,1:m})&:=p(\theta_0,\varphi_0|Y_{0,1:m}) \prod_{i=1}^n p(\theta_i|Y_{i,1:m},\varphi_0).
\end{aligned}
\label{E12}
\end{equation}
Here, estimation of $\varphi_0$ depends on only $Y_{0,1:m}$. Contributions from $Y_{1:n,1:m}$ at other locations are completely removed.

\subsection{Manipulating the multiple feedback and the Bayesian GWR posterior}
Suppose now that $\varphi(u,v)$ is not constant but is a smooth function with respect to $(u,v)$ so that closer locations have more similar $\varphi$. In this case it is inappropriate to treat the misspecification as equally problematic at every location since this may lead to a loss of usable information from the dataset. Instead we propose to manipulate contributions to the estimation of $\varphi_0$ from observations $Y_{i,1:m}$ neighbouring the location of interest $i= 0$ by varying amounts. We achieve this by allocating a geographically weighted kernel $W(d_i,\eta)$ to the likelihood of $Y_{i,1:m}$ where $d_i$ is the distance between location $0$ and location $i$. 

\begin{figure}[t] 
\setlength{\abovecaptionskip}{0cm}
\setlength{\belowcaptionskip}{0cm}
\centering 
\includegraphics[scale=0.5]{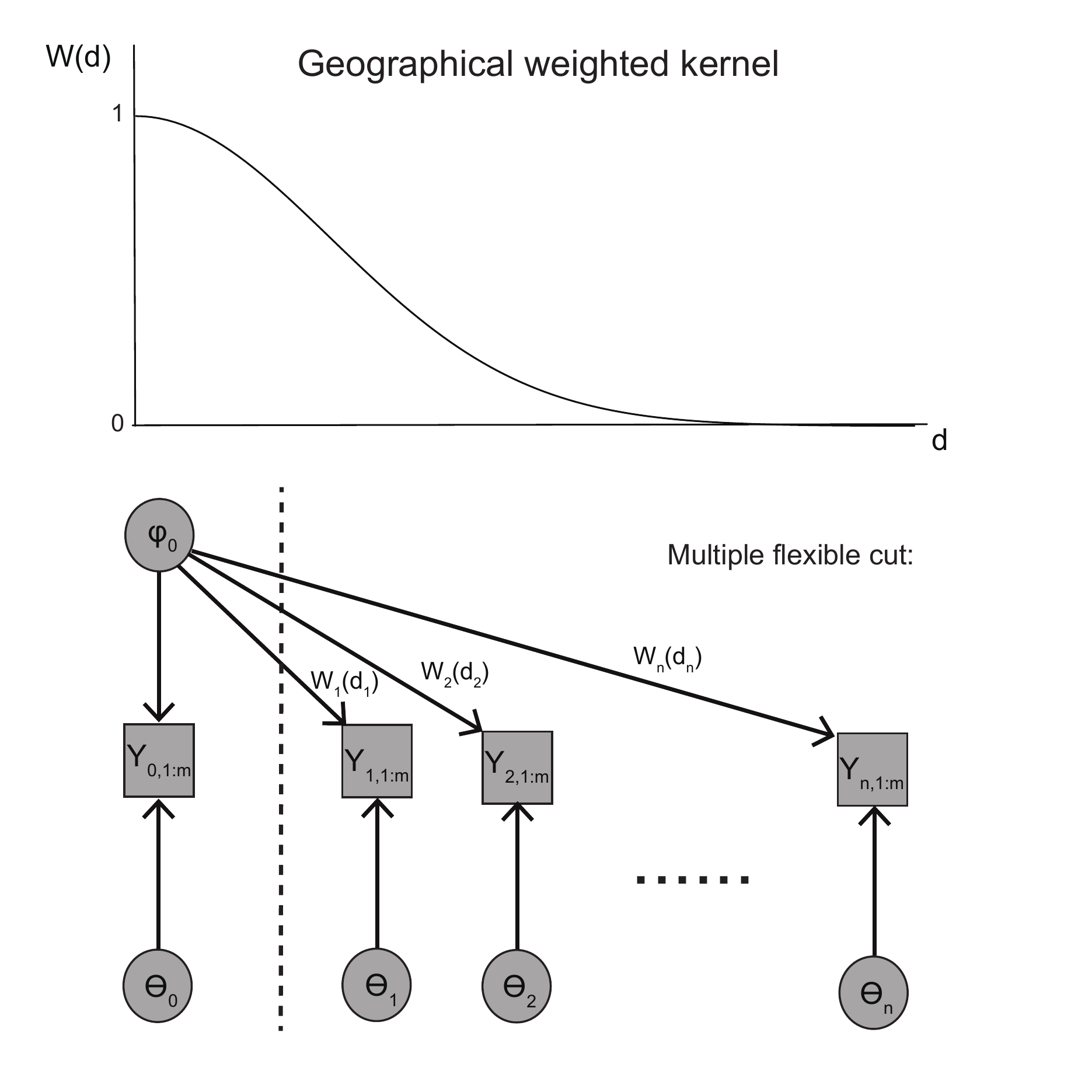} 
\caption[DAG representation when the feedback is manipulated.]{\textbf{DAG representation when the feedback is manipulated.} The $n+1$ modules $(Y_{i,1:m}, \varphi_0,\theta_i)$, $i=0,\dots,n$, are separated by a dashed line. The location of interest is $i=0$.} 
\label{F3}
\end{figure}

Figure \ref{F3} shows a DAG of this model. It can be viewed as a case of manipulating the feedback between $n+1$ modules. Extending \cite{carmona2020semi}, we introduce an auxiliary variable $\tilde{\theta}_{1:n}=(\tilde{\theta_1},\cdots,\tilde{\theta_n})$, which has the same likelihood term as $\theta_{1:n}$. We set an independent prior $\pi(\theta_0,\tilde{\theta}_{1:n},\varphi_0)=\prod_{i=1}^n \pi(\tilde{\theta}_i)\pi(\theta_0)\pi(\varphi_0)$. Then we write
\begin{equation}
p_\eta(\theta_{0:n},\tilde{\theta}_{1:n},\varphi_0|Y_{0:n,1:m})=p_{\text{pow},\eta}(\theta_0,\tilde{\theta}_{1:n},\varphi_0|Y_{0:n,1:m})\prod_{i=1}^n p(\theta_i|Y_{i,1:m},\varphi_0),
\label{E13}
\end{equation}
where
\begin{equation}
p_{\text{pow},\eta}(\theta_0,\tilde{\theta}_{1:n},\varphi_0|Y_{0:n,1:m})\propto p(Y_{0,1:m}|\theta_0,\varphi_0) \pi(\theta_0,\tilde{\theta}_{1:n},\varphi_0) \prod_{i=1}^n p(Y_{i,1:m}|\tilde{\theta}_i,\varphi_0)^{W(d_i,\eta)} 
\label{E14}
\end{equation}
is called the geographically-powered posterior and is used to adjust contributions from observations $Y_{i,1:m}$ by allocating the corresponding weighted kernel $W(d_i,\eta)$ to the likelihood $p(Y_{i,1:m}|\varphi_0,\tilde{\theta}_i)$. Note that \eqref{E14} is an extension of the usual power posterior and it contains the GWR locally-weighted likelihood \eqref{E3}. Given the geographical bandwidth $\eta$, the SMI distribution for this multiple module case is
\[
p_{\eta}(\theta_{0:n},\varphi_0|Y_{0:n,1:m})=\int p_\eta(\theta_{0:n},\tilde{\theta}_{1:n},\varphi_0|Y_{0:n,1:m}) d\tilde{\theta}_{1:n}.
\]
The Bayesian GWR posterior for the parameters of interest $\varphi_0$ and $\theta_0$ at the location of interest $i =0$ is
\begin{equation}
\begin{aligned}
p_{\eta}(\theta_0,\varphi_0|Y_{0:n,1:m})&=\int p_{\eta}(\theta_{0:n},\varphi_0|Y_{0:n,1:m}) d\theta_{1:n}\\
&=\int \int p_\eta(\theta_{0:n},\tilde{\theta}_{1:n},\varphi_0|Y_{0:n,1:m}) d\tilde{\theta}_{1:n}
d\theta_{1:n} \\
&=\int p_{\text{pow},\eta}(\theta_0,\tilde{\theta}_{1:n},\varphi_0|Y_{0:n,1:m}) d\tilde{\theta}_{1:n}.
\end{aligned}
\label{E16}
\end{equation}
We call estimation of the parameter of interest $(\theta_0,\varphi_0)$ via \eqref{E16} Bayesian GWR inference. The Bayesian GWR model manipulates the feedback from each of the multiple neighbouring observations through the geographical bandwidth $\eta$, and reduces to the cut distribution and the standard posterior distribution for certain values for $\eta$. Specifically, when the variation of $\varphi(u,v)$ is so large that we are not confident to include neighbouring locations, then $\eta\rightarrow 0$ and the estimation of $\theta_0$ and $\varphi_0$ only depends on observations $Y_{0,1:m}$.
\[
\lim_{\eta\rightarrow 0}p_{\eta}(\theta_{0:n},\varphi_0|Y_{0:n,1:m})=p(\theta_0,\varphi_0|Y_{0,1:m}) \prod_{i=1}^n p(\theta_i|Y_{i,1:m},\varphi_0) = p_{\text{cut}}(\theta_{0:n},\varphi_0|Y_{0:n,1:m}).
\]
This is the cut distribution \eqref{E12}. In contrast, when the variation of $\varphi(u,v)$ is so small that we can include observations from all locations, then $\eta\rightarrow \infty$ and estimation of $\theta_0$ and $\varphi_0$ depends on all observations $Y_{0:n,1:m}$ as in the standard posterior distribution \eqref{E10}:
\[
\lim_{\eta\rightarrow \infty}p_{\eta}(\theta_{0:n},\varphi_0|Y_{0:n,1:m})=p(\theta_0,\varphi_0|Y_{0:n,1:m}) \prod_{i=1}^n p(\theta_i|Y_{i,1:m},\varphi_0) = p(\theta_{0:n},\varphi_0|Y_{0:n,1:m}).
\]
In summary, we propose the Bayesian GWR model for multiple suspect modules for the situation that the geographical weighted kernel \eqref{E4} has a known and deterministic functional form with respect to the geographical coordinates. Since the joint `likelihood' involved in \eqref{E14} is the geographically weighted likelihood widely used in the GWR framework, the essence of the Bayesian GWR model is a particular extension of the SMI model.

\subsection{Theoretical analysis}
Bayes' theorem can not be used to justify the proposed geographically-powered posterior because the power likelihood is not a proper probability distribution. Instead we justify the geographically-powered posterior as a minimizing rule within an information processing framework, thus avoiding the need to appeal to Bayes' theorem. We also study its property subject to a large sample size.

We write the true data generating process for the complete set of observations $Y_{0:n,1:m}$ as
\[
\check{p}_{0:n,1:m}(Y_{0:n,1:m})=\prod_{i=0}^n \check{p}_{i,1:m}(Y_{i,1:m})=\prod_{i=0}^n \prod_{j=1}^m \check{p}_i(Y_{i,j}),
\]
where $\check{p}_i$ is the true generating process at location $i$. Let $\check{P}_{0:n,1:m}$ be the corresponding probability measure. Denoting $\psi=(\theta_0,\tilde{\theta}_{1:n},\varphi_0) \in \Psi$ and $W_i=W(d_i,\eta)$ and omitting $\eta$ in $p_{\text{pow},\eta}$ for simplicity, the geographically-powered likelihood $p_{\text{pow}}(Y_{0:n,1:m}|\psi)$ for observations $Y_{0:n,1:m}$ is written as \eqref{E3} where $\theta_i$ is replaced with $\tilde{\theta}_i$ for $i\neq 0$. Let $\Pi$ be the probability measure of prior distribution. If $p_{\text{pow}}(Y_{0:n,1:m}):=\int p_{\text{pow}}(Y_{0:n,1:m}|\psi) \Pi(d\psi)<\infty$, we can re-write the probability measure of geographically-powered posterior \eqref{E14} on any $\Psi^\ast\subset\Psi$ in terms of the true data generating processes as
\begin{equation}
\begin{aligned}
P_{\text{pow}}(\Psi^\ast|Y_{0:n,1:m}) = \frac{\int_{\Psi^\ast} \exp\left\{-r_{0,1:m}(\psi)-\sum_{i=1}^n W_ir_{i,1:m}(\psi)\right\} \Pi(d\psi)}{\int_{\Psi} \exp\left\{-r_{0,1:m}(\psi)-\sum_{i=1}^n W_ir_{i,1:m}(\psi)\right\} \Pi(d\psi)},
\end{aligned}
\label{SE1}
\end{equation}
where
\[
r_{0,1:m}(\psi) = \log\left\{\frac{\check{p}_{0,1:m}(Y_{0,1:m})}{p(Y_{0,1:m}|\theta_0,\varphi_0)}\right\};\ r_{i,1:m}(\psi) = \log\left\{\frac{\check{p}_{i,1:m}(Y_{i,1:m})}{p(Y_{i,1:m}|\tilde{\theta}_i,\varphi_0)}\right\}\ i\neq 0.
\]
This representation makes it clear that \eqref{SE1} is an extension of the Gibbs posterior \citep{jiang2008}, which is also known as the generalized Bayesian posterior \citep{grunwald2017}; pseudo posterior \citep{https://doi.org/10.1111/1467-9868.00314,JMLR:v17:15-290}; and quasi-posterior \citep{CHERNOZHUKOV2003293,doi:10.1080/10485250500039049}), which plays an essential role in the study of the PAC-Bayesian inference \citep[e.g.,][]{dalalyan2008aggregation,LEVER20134}. The Gibbs posterior generalizes the usual Bayesian posterior by defining a prior for the parameter of a loss function, which need not be the negative log-likelihood as used in standard Bayesian inference.

Our model extends the existing Gibbs posterior literature by allowing multiple learning rates (also interpreted as temperatures in thermodynamics \citep{4767596}) which correspond to geographically weighted kernels. The loss function (or the statistical risk function) at each location is $W_i r_{i,1:m}(\psi)$, where $W_0=1$. We denote the empirical total loss function $L_{1:m}(\psi)$, given the parameter of the model $\psi$, as:
\[
L_{1:m}(\psi) = \frac{1}{m}\left(r_{0,1:m}(\psi)+\sum_{i=1}^n W_i r_{i,1:m}(\psi)\right). 
\]

%abandon
\if 0
\[
\begin{aligned}
L_{1:m}(\psi) &= \frac{1}{m}\left(r_{0,1:m}(\psi)+\sum_{i=1}^n W_i r_{i,1:m}(\psi)\right) \\
&= \frac{1}{m}\left(\sum_{j=1}^m \left\lbrace \log\left(\frac{\check{p}_0(Y_{i,j})}{p(Y_{i,j}|\theta_0,\varphi_0)}\right)+\sum_{i=1}^n W_i \log\left(\frac{\check{p}_i(Y_{i,j})}{p(Y_{i,j}|\tilde{\theta}_i,\varphi_0)}\right) \right\rbrace \right),
\end{aligned}
\]
\fi
%abandon

Let $F$ be a probability measure on the parameter space $\Psi$ which results from processing the information from observations $Y_{0:n,1:m}$ and prior knowledge $\Pi$. We aim to show that the geographically-powered posterior $P_{\text{pow}}$ is the optimal $F$ in the sense that $P_{\text{pow}}$ minimizes an information bound. We first need to construct this information bound. \cite{bhattacharya2019} provides a PAC-Bayesian type bound for the power posterior. The bound controls a R\' enyi divergence which characterizes the performance of the power posterior. We now denote the R\' enyi divergence between two arbitrary distribution $p$ and $q$, given an $\alpha\in (0,1)$, as:
\[
\mathbb{D}_\alpha(p(\cdot),q(\cdot)) = \frac{1}{\alpha-1}\log\left(\int p(y)^\alpha q(y)^{1-\alpha} dy\right).
\]
We have the following theorem that extends the Theorem 3.4 of \cite{bhattacharya2019} by allowing multiple learning rates.

\begin{theorem}[Weighted R\' enyi divergence bound]
\label{THE1}
Given a distribution $f(\psi)$ with probability measure $F(\cdot)$ over parameter space $\Psi$, for any $\varepsilon\in (0,1)$, the following inequality
\[
\begin{aligned}
&\int \sum_{i=1}^n (1-W_i) \mathbb{D}_{W_i}\left(p(\cdot|\tilde{\theta}_i,\varphi_0),\check{p}_i(\cdot)\right)F(d\psi) \\
& \ \ \leq \frac{1}{m}\int \left(r_{0,1:m}(\psi) +\sum_{i=1}^n W_i r_{i,1:m}(\psi)\right)F(d\psi) + \frac{\mathbb{D}_{KL}(f(\cdot),\pi(\cdot))}{m}+\frac{1}{m}\log\left(\frac{1}{\varepsilon}\right)
\end{aligned}
\]
holds with $\check{P}_{0:n,1:m}$ probability at least $(1-\varepsilon)$.
\end{theorem}
\begin{proof}
See appendix.
\end{proof}

\begin{remark}
Theorem \ref{THE1} leads to the following ``information posterior bound'' \citep{1614067}, which holds with $\check{P}_{0:n,1:m}$ probability at least $(1-\varepsilon)$.
\[
\resizebox{\textwidth}{!}{$\underset{\psi\sim F}{\mathbb{E}} \left\{ -\log\left(\underset{Y_{0:n,1:m}\sim \check{P}_{0:n,1:m}}{\mathbb{E}} \exp\left(- L_{1:m}(\psi)\right)\right) \right\} \leq \underset{\psi\sim F}{\mathbb{E}} L_{1:m}(\psi) +\frac{\mathbb{D}_{KL}(f(\cdot),\pi(\cdot))}{m}+\frac{1}{m}\log\left(\frac{1}{\varepsilon}\right)$}
\]
For a proof, see supplementary materials.
\end{remark}

Given a distribution $F$ which results from an information processing rule, the Remark states that the negative logarithm of the expected exponential of the negative loss is controlled by the empirical loss from the usage of $F$ and an additional penalty on the discrepancy between $F$ and the prior $\Pi$. \cite{1614067} proposed an approach called ``Information Risk Minimization'' which selects $F$ by minimizing the right hand side of the information posterior bound. Note that, although the bound involves $\varepsilon$, the inequality holds for any $\varepsilon\in (0,1)$. Hence, the selection of $F$ is not affected by $\varepsilon$. Similarly, the true data generating process drops out since it does not involve $F$. To apply this approach, it is equivalent to find a $F$ that minimizes the following criterion function
\begin{multline*}
M_m(f(\psi)) = \mathbb{D}_{KL}(f(\cdot),\pi(\cdot)) - \int f(\psi) \log\left(p(Y_{0,1:m}|\theta_0,\varphi_0)\right) d\psi\\
-\sum_{i=1}^n W_i \int f(\psi) \log\left(p(Y_{i,1:m}|\tilde{\theta}_i,\varphi_0)\right)  d\psi.
\end{multline*}
Note that the ``Information Risk Minimization'' used here can be regarded as a modified ``Information Conservation Principle'' \citep{doi:10.1080/00031305.1988.10475585}. This principle states that an optimal information processing rule has equal input information $I_{\text{in}}$, which consists the information processing (i.e., prior knowledge, observations and model), and output information $I_{\text{out}}$. In our setting, for the probability measure $F$, the input information $I_{\text{in}}$ is:
\[
\begin{aligned}
I_{\text{in}} &:=  \int \log(\pi(\psi)) F(d\psi) + \int  \log\left(p_{\text{pow}}(Y_{0:n,1:m}|\psi)\right) F(d\psi) \\
&= \int \log(\pi(\psi)) F(d\psi) + \int \log\left(p(Y_{0,1:m}|\theta_0,\varphi_0)\right) F(d\psi) \\
& \quad + W_i \sum_{i=1}^n \int \log\left(p(Y_{i,1:m}|\tilde{\theta}_i,\varphi_0)\right) F(d\psi).
\end{aligned}
\]
Note that in contrast to the original input information discussed in \cite{doi:10.1080/00031305.1988.10475585}, the input information from each geographical location is manipulated by the geographically weighted kernel. The output information $I_{\text{out}}$ is:
\[
I_{\text{out}} := \int \log(f(\psi)) F(d\psi) + \int \log(p_{\text{pow}}(Y_{0:n,1:m})) F(d\psi).
\]
Now we present the following theorem which justifies the use of the geographically-powered posterior \eqref{SE1} as the form of probability distribution that statistically learns information from the observations and the prior knowledge while minimising the loss.

\begin{theorem}[Justification]
\label{THE2}
If $p_{\text{pow}}(Y_{0:n,1:m}):=\int p(Y_{0:n,1:m}|\psi)\pi(\psi)d\psi <\infty$, the geographically-powered posterior $p_{\text{pow}}(\psi|Y_{0:n,1:m})$ minimizes the criterion function $M_m(f(\psi))$ with respect to a probability distribution $f(\psi)$. In addition, the geographically-powered posterior results from the optimal information processing rule.
\end{theorem}
\begin{proof}
See appendix.
\end{proof}

We now consider the large sample size setting. Let $y_{0:n}=\{y_0,y_1,\cdots,y_n\}$ be random variables corresponding to a single observation at each location and $y_{0:n}\sim \check{P}=\prod_{i=1}^n \check{P}_i$. Although the GWR model is less necessary in the large sample size setting (since effective statistical inference can be conducted separately at each location), we wish to show that the posterior predictive distribution $p(y_i|\tilde{\theta}_i,\varphi_0)$, where $(\tilde{\theta}_i,\varphi_0)\sim P_{\text{pow}}$, approaches the truth $\check{p}_i$ at each location $i=0,1,\cdots,n$ when the degree of the partial misspecification varies across the geographical space. Denote the expected total loss function $L(\psi)$, given the parameter of the model $\psi$, as:
\[
L(\psi) = \underset{y_{0:n}\sim \check{P}}{\mathbb{E}} \left\lbrace \log\left(\frac{\check{p}_0(y_0)}{p(y_0|\theta_0,\varphi_0)}\right) + \sum_{i=1}^n W_i \log\left(\frac{\check{p}_i(y_i)}{p(y_i|\tilde{\theta}_i,\varphi_0)}\right) \right\rbrace.
\] 
We present the following theorem.

\begin{theorem}[Consistency]
\label{THE3}
Given a finite number of observations, the geographically-powered posterior $P_{\text{pow}}$ minimizes
\[
\underset{\psi\sim P_{\text{pow}}}{\mathbb{E}}\left( L_{1:m}(\psi)\right) + m^{-1}\mathbb{D}_{KL}(p_{\text{pow}}(\cdot|Y_{0:n,1:m}),\pi(\cdot)).
\]
When the sample size $m \rightarrow \infty$ at all locations and suppose that the limit of the geographically-powered posterior $P_{\text{pow}}^{(\infty)}:=\lim_{m \rightarrow \infty}P_{\text{pow}}$ exists, then $P_{\text{pow}}^{(\infty)}$ puts all its mass at $\psi^\ast=(\theta_0^\ast,\tilde{\theta}_{1:n}^\ast,\varphi_0^\ast)$ which minimizes the expected total loss function (a geographically weighted combination of Kullback-Leibler divergences):
\[
\begin{aligned}
\psi^\ast &= \argmin_{\psi=(\theta_0,\tilde{\theta}_{1:n},\varphi_0)} L(\psi) \\
&= \argmin_{\psi=(\theta_0,\tilde{\theta}_{1:n},\varphi_0)} \mathbb{D}_{KL}\left(\check{p}_0(\cdot),p(\cdot|\theta_0,\varphi_0)\right)+\sum_{i=1}^n W_i \mathbb{D}_{KL}\left(\check{p}_i(\cdot),p(\cdot|\tilde{\theta}_i,\varphi_0)\right).
\end{aligned}
\]
\end{theorem}
\begin{proof}
See appendix.
\end{proof}

Although partial misspecification remains and predictions drawn from the model will not follow the true data generating process, Theorem \ref{THE3} states that the geographically-powered posterior draws predictions that balance minimizing the empirical total loss function and the discrepancy between posterior and prior knowledge. When the sample size increases, the model acts similarly to a standard Bayesian model by learning more from observations. In the limit of an infinite sample size, the model provides a prediction that is closest to the true data generating process. Note that, although the model draws predictions close to the truth, more priority is assigned to locations close to the location of interest, and so we cannot use a single Bayesian GWR model when inference is needed for multiple locations. Instead, separate models should be used at each location of interest.

\section{Inference for Multiple Locations and Bandwidth Selection}
\subsection{Predictive performance of one Bayesian GWR model}
In Section \ref{SEC3}, we considered the setting when there is a single location of interest. We now consider inference for multiple sampling locations when all locations are of interest. This is done by using separate Bayesian GWR models for each location while assuming the same geographical bandwidth for all models. We give the following definition which generalizes the Bayesian GWR model by relaxing the location of interest.

\begin{definition}
Consider observations $Y_{i,1:m}$ sampled from location $i$ with coordinate $(u_i,v_i)$, $i=0,\cdots,n$; a bandwidth $\eta$ and a specific geographical coordinate $(u,v)$ that we call the geographical centre. Define the Bayesian GWR model $M=((u,v),\eta)$ with parameter $\psi_M=(\theta_{M,0:n},\varphi_M)$ to be the SMI model with distribution
\begin{equation}
    p_{M}(\psi_M|Y_{0:n,1:m})=\int p_M(\psi_M,\tilde{\theta}_{M,0:n}|Y_{0:n,1:m}) d\tilde{\theta}_{M,0:n},
\label{E20}
\end{equation}
where $\tilde{\theta}_{M,1:n}$ is the auxiliary variable for model $M$ and
\begin{equation}
\begin{split}
p_M(\psi_M,\tilde{\theta}_{M,0:n}|Y_{0:n,1:m}) &\propto \pi(\tilde{\theta}_{M,0:n},\varphi_M)\\
& \quad \times \prod_{i=0}^n p(Y_{i,1:m}|\tilde{\theta}_{M,i},\varphi_M)^{W(d_i,\eta)} \prod_{i=0}^n p(\theta_{M,i}|Y_{i,1:m},\varphi_M),
\label{E21}
\end{split}
\end{equation}
where $d_i$ is the geographical distance between location $(u_i,v_i)$ and the geographical centre $(u,v)$. In the special case when the geographical centre is one of the sampling locations, which we assume without loss of generality to be $(u_0,v_0)$, then \eqref{E20} and \eqref{E21} reduce to  \eqref{E13} and \eqref{E14}.
\end{definition}

To measure the predictive performance of a model $M$ for, for example, a new observation $Y_i^\ast$ from location $i$ with true generating process $\check{p}_i(Y_i^\ast)$, we use the Kullback-Leibler (KL) divergence. This is achieved by looking at the expected log pointwise predictive density \citep{gelman2014understanding, jacob2017better}, which is essentially a constant term minus the KL divergence, and is defined as
\begin{equation}
\text{elpd}_{(u_i,v_i)}(M) := \int \check{p}_i(Y_i^\ast) \log \left(p_{M}(Y_i^\ast|Y_{0:n,1:m})\right) dY_i^\ast,
\label{E29}
\end{equation}
where the predictive distribution $p_{M}(Y_i^\ast|Y_{0:n,1:m})$ is defined as
\[
\begin{aligned}
    p_{M}(Y_i^\ast|Y_{0:n,1:m}) &:=\int p(Y_i^\ast|\theta_{M,i},\varphi_M)p_{M}(\psi_M|Y_{0:n,1:m}) d\psi_M \\
    &= \int p(Y_i^\ast|\theta_{M,i},\varphi_M)p_{M}(\theta_{M,i},\varphi_M|Y_{0:n,1:m}) d\theta_{M,i}d\varphi_M.
\end{aligned}
\]
Here, we denote $p_{M}(\theta_{M,i},\varphi_M|Y_{0:n,1:m}):=\int p_{M}(\psi_M|Y_{0:n,1:m}) d\theta_{M,-i}$, where we define $\theta_{M,-i}=(\theta_{M,0},\cdots,\theta_{M,i-1},\theta_{M,i+1},\cdots,\theta_{M,n})$.

\subsection{Inference for multiple locations}
Having defined the measure of predictive performance for one Bayesian GWR model, we are ready to extend it to infer multiple locations by setting and tuning multiple Bayesian GWR models. The following assumption can be viewed as a rephrasing of the first law of geography \citep{tobler1970computer}, since for an arbitrary location of interest $i$, observations from closer locations contribute more to the estimation of the shared parameter $\varphi$ when the geographical centre is exactly equal to the location of interest.
\begin{assumption}
\label{ASS1}
For any fixed geographical bandwidth $\eta$ and specific location with geographical coordinates $(u_k,v_k)$, $\text{elpd}_{(u_k,v_k)}(M)$ is maximized when the geographical centre $(u,v)=(u_k,v_k)$. That is:
\[
    ((u_k,v_k),\eta)=\argmax_M \text{elpd}_{(u_k,v_k)}(M),\ \forall \eta.
\]
\end{assumption}
We define the space of Bayesian GWR models $\mathcal{M}=\{M=((u,v),\eta): \eta>0\}$. The following assumption assumes inferences from multiple models are independent.
\begin{assumption}
\label{ASS2}
Given a dataset $Y_{0:n,1:m}$ and Bayesian GWR models $M_s\in\mathcal{M}$, $s=1,\cdots,S$, we have the joint Bayesian GWR posterior
\[
    p(\psi_{M_1},\cdots,\psi_{M_S}|Y_{0:n,1:m})=\prod_{s=1}^S p_{M_s}(\psi_{M_s}|Y_{0:n,1:m}).
\]
\end{assumption}

We are now ready to extend inference to multiple locations. Given a set of Bayesian GWR models $M=(M_0,\cdots,M_n)$, one for each geographic sampling location, all with identical geographical bandwidth $\eta$, we define the expected log pointwise predictive density for new observations $Y_{0:n}^\ast=(Y_0^\ast,\cdots,Y_n^\ast)$ with each single observation $Y_i^\ast$ from location $i$ as
\[
    \text{elpd}(M) = \int \log \left(p(Y_{0:n}^\ast|Y_{0:n,1:m})\right) \prod_{i=0}^n \check{p}_i(Y_i^\ast)dY_{0:n}^\ast,
\]
where
\[
    p(Y_{0:n}^\ast|Y_{0:n,1:m}) = \int p(Y_{0:n}^\ast|\psi_{M_0},\cdots,\psi_{M_n}) p(\psi_{M_0},\cdots,\psi_{M_n}|Y_{0:n,1:m}) d\psi_{M_0}\cdots d\psi_{M_n}.
\]
We then present the following theorem to select the optimal bandwidth.

\begin{theorem}[Bandwidth selection]
\label{THE5}
Given Assumption \ref{ASS1} and \ref{ASS2}, for observations $Y_{0:n,1:m}$ sampled from locations $i$ with coordinates $(u_i,v_i)$, $i=0,\cdots,n$, the optimal combination of $n+1$ separate Bayesian GWR models $M=(M_0,\cdots,M_n)$ that maximizes $\text{elpd}(M)$, where each $M_i=((u_i^\ast,v_i^\ast),\eta^\ast)$ is used for prediction in location $i$, satisfies
\begin{enumerate}
    \item For all $0\leq i \leq n$,
    \[
    (u_i^\ast,v_i^\ast)=(u_i,v_i).
    \]

    \item Redefine $M_i(\eta)=((u_i,v_i),\eta)$, then the optimal bandwidth $\eta^\ast$ maximizes the mean (across all sampling locations) expected log pointwise predictive density.
    \[
    \eta^\ast = \argmax_\eta \frac{1}{n+1} \sum_{i=0}^n \text{elpd}_{(u_i,v_i)} (M_i(\eta)) .
    \]
\end{enumerate}
\end{theorem}
\begin{proof}
See appendix.
\end{proof}

% abandoned
\if 0
Theorem \ref{THE5} indicates that the optimal bandwidth $\eta$ is selected that minimizes the average KL divergence between the true data generating process $\check{p}_i(y_i)$ and the modeled generating process $p_{M_i}(y_i|Y_{0:n,1:m})$ across all sampling locations. In addition, each modeled generating process $p_{M_i}(y_i|Y_{0:n,1:m})$ is built by allocating the geographical centre to be exactly equal to the location of interest (discussed in Section 2.2).
\fi

In practice, we do not know the true data generating process $\check{p}_i$. Numerous methods \citep[e.g.,][]{gelman2014understanding} can be applied to approximate \eqref{E29}. Here, we adopt cross-validation to estimate $\text{elpd}_{(u_i,v_i)} (M_i)$ because it measures out-of-sample predictive performance and consequently avoids overestimating elpd. We train the model $M_i$ on all observations from other locations $Y_{j,1:m}$, $j\neq i$ and a subset $Y_{i,1:m^\prime}$ of the observations from location $i$ (denoted as $\{Y_{0:n,1:m}\setminus Y_{i,m^\prime+1:m}\}$), and estimate elpd using the test set $Y_{i,m^\prime+1:m}$ by
\begin{equation}
\widehat{\text{elpd}}_{(u_i,v_i)} (M_i)= \frac{1}{m-m^\prime}\sum_{j=m^\prime+1}^m \log\left(\int p(Y_{i,j}|\psi_{M_i})p_{M_i}(\psi_{M_i}|\{Y_{0:n,1:m}\setminus Y_{i,m^\prime+1:m}\})d\psi_{M_i}\right).
\label{E30}
\end{equation}
The integral within \eqref{E30} can be easily approximated by the Monte Carlo samples drawn from the Bayesian GWR posterior. This is summarized in Algorithm \ref{Al1}. 

\subsection{Algorithm and simplification of computation}
We summarize the algorithm for the Bayesian GWR model when there are $n+1$ locations. For a set of candidate geographical bandwidths $\{\eta_r\}_{r=1}^R$, we select the optimal geographical bandwidth $\eta$ using Algorithm \ref{Al1}. In Algorithm \ref{Al2}, samples at each iteration can be drawn by using any standard sampler (e.g., Metropolis-Hastings or Gibbs sampler). The algorithm requires an approximation of the elpd at each location separately. This can be done in parallel to expedite computation. Once the optimal geographical bandwidth $\eta$ has been selected, we refit model with this bandwidth to the whole dataset, as described in Algorithm \ref{Al2}. We provide the code for both algorithms in Python Version 3 (\url{https://github.com/MathBilibili/Bayesian-geographically-weighted-regression}).

\begin{algorithm}[!htb] 
\caption{Selection of geographical bandwidth $\eta$ by cross-validation}\label{Al1}
\begin{algorithmic}[1]
\Require{A candidate set of geographical bandwidths $\{\eta_r\}_{r=1}^R$, observations $Y_{0:n,1:m}$ and its corresponding coordinates $\{(u_i,v_i)\}_{i=0}^n$, likelihood $p(Y|\theta,\varphi)$, prior $\pi(\theta)$, $\pi(\tilde{\theta})$ and $\pi(\varphi)$, number of iterations $S$, number $Q$ of $k$-fold cross-validation folds.} 
\Statex
 \For{$r \in \{1,\cdots,R\}$}
  \For{$q \in \{1,\cdots,Q\}$}  
  \State 
  \parbox[t]{\dimexpr\linewidth-3.1em}{Select the test set $Y_{i,(m^\prime+1:m)}$, a random $100/k\%$ subset of observations at location $i$, and training set $Y^{(i)}= Y_{0:n,1:m} \setminus Y_{i,(m^\prime+1:m)}$ for location $i$, $i=0, \dots, n$.}
  \State 
  \parbox[t]{\dimexpr\linewidth-3.1em}{Call Algorithm \ref{Al2} with Bayesian GWR models $M_i(\eta_r)=((u_i,v_i),\eta_r)$ and location-specific dataset $Y^{(i)}$, $i=0, \dots, n$.}
    \State 
    \parbox[t]{\dimexpr\linewidth-3.1em}{Calculate  $\widehat{\text{elpd}}_{(u_i,v_i)}(M_i(\eta_r))$ on the test set $Y_{i,(m^\prime+1:m)}$ using samples $\{(\varphi_i^{(s)},\theta_i^{(s)})\}_{s=1}^{S}$, $i=0, \dots, n$.\strut}
   \State Calculate $q^{th}$ mean elpd: $\overline{\text{elpd}_q}(\eta_r) = \frac{1}{n+1}\sum_{i=0}^n\widehat{\text{elpd}}_{(u_i,v_i)}(M_i(\eta_r))$.
      \EndFor
  \EndFor
    \State \Return {$\{\{\overline{\text{elpd}_q}(\eta_r)\}_{q=1}^Q\}_{r=1}^R$.}
\end{algorithmic}
\end{algorithm}

\begin{algorithm}[!htb] 
\caption{Bayesian GWR model for multiple locations}\label{Al2}
\begin{algorithmic}[1]
\Require{A geographical bandwidth $\eta$, observations $Y_{0:n,1:m}$ and corresponding coordinates $\{(u_i,v_i)\}_{i=0}^n$, likelihood $p(Y|\theta,\varphi)$, prior $\pi(\theta)$, $\pi(\tilde{\theta})$ and $\pi(\varphi)$, number of iterations $S$.} 
\Statex 
\State Set Bayesian GWR models $M_i(\eta)=((u_i,v_i),\eta)$ and location-specific dataset $Y^{(i)}$, $i=0, \dots, n$. Note that $Y^{(i)}=Y_{0:n,1:m}$ if cross-validation is not required.
  \For{$i \in \{0,\cdots,n\}$}
   \State 
   \parbox[t]{\dimexpr\linewidth-3.2em}{Calculate geographically weighted kernels, where the distance is calculated between $(u_j,v_j)$ and geographical centre $(u_i,v_i)$ for $j=0,\dots,n$.\strut}
    \State 
    \parbox[t]{\dimexpr\linewidth-3.2em}{Draw samples $\{\theta_i^{(s)},\tilde{\theta}_{-i}^{(s)},\varphi_i^{(s)}\}$ from $p_{\text{pow},\eta}(\theta_i,\tilde{\theta}_{-i}, \varphi_i|Y^{(i)})$, $s=1, \dots, S$, according to \eqref{E14}, with location of interest \(i\) and $\tilde{\theta}_{-i}=(\tilde{\theta}_0,\cdots,\tilde{\theta}_{i-1},\tilde{\theta}_{i+1},\cdots,\tilde{\theta}_n)$.\strut}
  \EndFor
    \State \Return {Bayesian GWR posterior samples $\{\{(\varphi_i^{(s)},\theta_i^{(s)})\}_{s=1}^S\}_{i=0}^n$.}
\end{algorithmic}
\end{algorithm}

The computational cost of a Bayesian GWR model for multiple locations is mainly determined by two factors when using a Metropolis-Hasting sampler. The first factor is the number of observations $m$ at each location, which clearly determines the number of likelihood evaluations required. In practice, this evaluation normally benefits from vectorization. 

The other factor is the number of locations $n$. On the one hand, by Assumption \ref{ASS2}, inference of parameters at each location is conducted using $n$ separate Bayesian GWR models, which can be easily parallelized. This can greatly reduce the computation time. On the other hand, when using the geographically weighted kernel \eqref{E4}, \eqref{E14} requires the powered likelihood to be evaluated $n$ times. When this computational cost is too large, it is possible to reduce the load by disregarding distant locations with only tiny weights. Specifically, inspired by the bi-square weighting function \citep{doi:10.1111/j.1538-4632.1996.tb00936.x}, a modified truncated Gaussian kernel may be useful:
\begin{equation}
W(d_i,\eta) = 
\begin{cases}
      \exp{\left(-\frac{d_i^2}{\eta^2}\right)} & \text{if $\exp{\left(-\frac{d_i^2}{\eta^2}\right)}>W^\ast$}\\
      0 & \text{otherwise}
    \end{cases}, 
\label{E31}
\end{equation}
where $W^\ast$ (e.g., $10^{-2}$) is a threshold value that controls the degree of exclusion. We want this exclusion to reduce the number of likelihood evaluations needed, while retaining all information from the neighbouring locations. A practical way to check this is by looking at the percentage change of the value of \eqref{E14} between kernels \eqref{E4} and \eqref{E31}. If the percentage change is trivial, \eqref{E31} will closely approximate \eqref{E4} but at much lower computational cost, especially when a small bandwidth $\eta$ is adopted. In summary when adopting kernel \eqref{E31}, the computational complexity, in terms of evaluating the likelihood of one observation of one MCMC iteration for one location of interest, is $\mathcal{O}(m \times n(W^\ast))$, where $n(W^\ast)$ is the number of locations for evaluations in \eqref{E14} with threshold $W^\ast$. 

\begin{figure}[h] 
\setlength{\abovecaptionskip}{0cm}
\setlength{\belowcaptionskip}{0cm}
\centering 
\includegraphics[width=.84\textwidth]{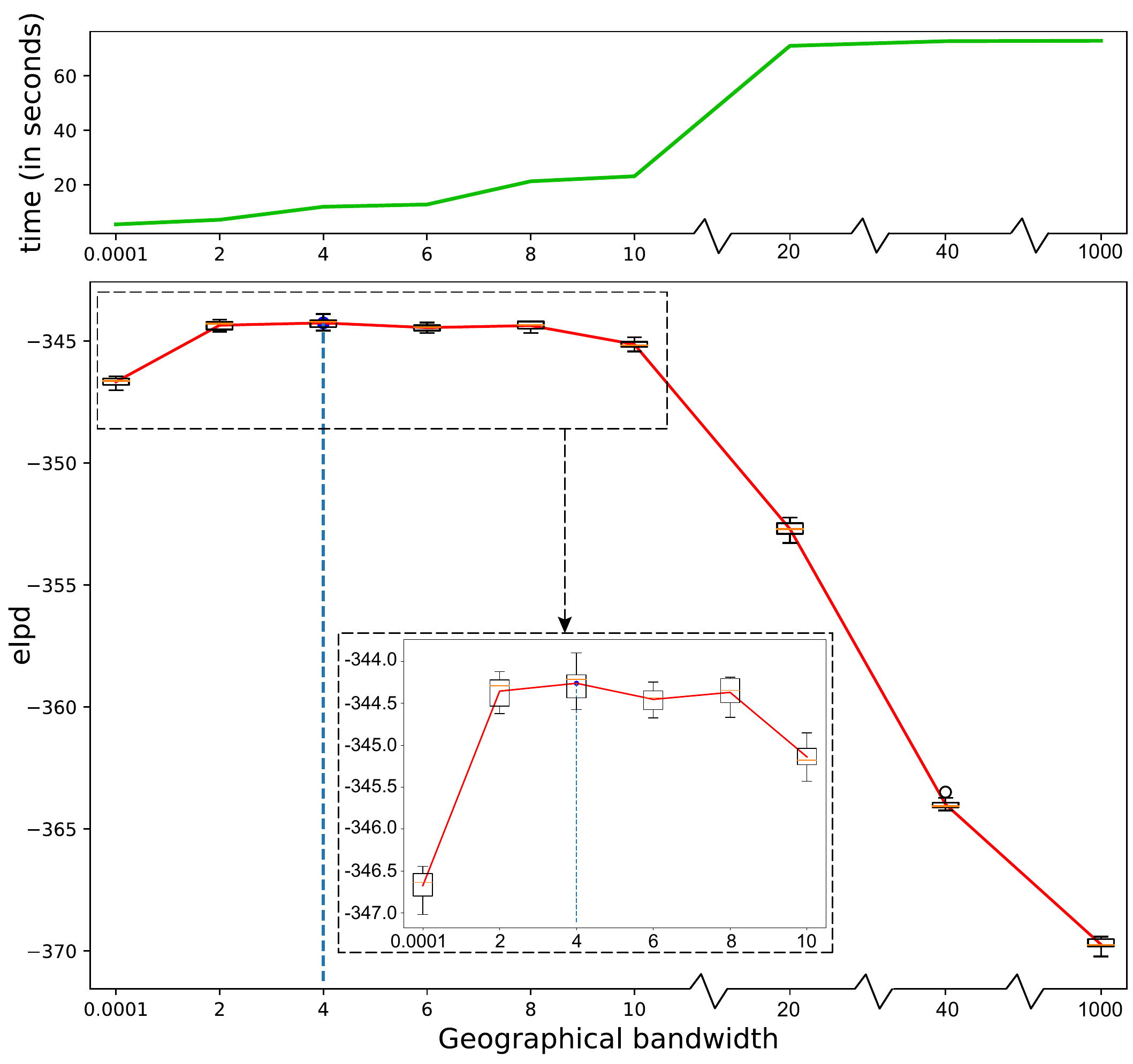} 
\caption[Computational time and elpd against geographical bandwidth.]{\textbf{Computational time and elpd against geographical bandwidth.} The computational time is calculated based on one MCMC iteration of running Bayesian GWR model for all locations. This is processed in parallel on ten cores of Intel Xeon E7-8860 v3 CPU. Each boxplot represents the elpd estimates from 10 chains across the whole geographic space. The red line is the average elpd estimates across the 10 chains. The blue dashed line indicates the optimal bandwidth. Two black dashed areas are equivalent with one being the other's expanding for a clear visualization.} 
\label{F4}
\end{figure}

\section{Simulation}
\label{SE5}
To illustrate our methodology and the influence of the geographical bandwidth, we simulated data on a $40\times 40$ regular lattice $(u,v)$, with $u=1,\cdots,40$ and $v=1,\cdots,40$, with geographically varying coefficients $\varphi=(\varphi_0,\varphi_1(u,v),\varphi_2(u))$ defined as:
\[
\begin{aligned}
\varphi_0&=3 \\
\varphi_1(u,v)&=0.1+0.01 \sqrt{u^2+v^2} \\
\varphi_2(u)&=0.05(\sin(\pi/2+\pi(u/20))+\cos(\pi/2+\pi(u/20))+4).
\end{aligned}
\]
We generated the true $\theta(u,v) \sim N(0.5, 0.01^2)$ independently: the resulting $\theta(u,v)$ is relatively constant across spatial locations, and its variability is not spatially smooth. With these coefficients, we simulated 100 independent samples at each location from a negative binomial distribution, with covariates $X=(X_1,X_2,X_3)$ where $X_1=1$ and $X_2$ and $X_3$ drawn from a uniform distribution $\textbf{U}(0,10)$ and $\textbf{U}(2,7)$.

We then fitted our Bayesian GWR model to each location separately and independently using the truncated Gaussian kernel with threshold $10^{-2}$, with geographical bandwidth $\eta$. The difference in \eqref{E14} using a truncated and non-truncated Gaussian kernel was less than $10^{-5}$\%, suggesting the truncated kernel closely approximates the non-truncated kernel. To estimate the elpd by cross validation, we excluded half of the samples at the location of interest from the training set. We drew $4\times10^3$ iterations for each of 10 independent chains at each location, discarding the first $1\times10^3$ samples as burn-in.

To identify the optimal geographical bandwidth $\eta$, we repeated this process for each of the 9 candidate values $\eta = 0.0001$, 2, 4, 6, 8, 10, 20, 40, and 1000. Figure \ref{F4} shows the computational time and estimated mean expected log pointwise predictive density (mean elpd across space), according to \eqref{E29}, for each candidate value. It can be seen that the mean elpd achieves its highest value when the bandwidth is 4, so we will compare results with $\eta=4$, $\eta=0.0001$ (the smallest candidate, equivalent to using samples only from the geographic centre) and $\eta=1000$ (the largest candidate, assuming the least geographic variation).

\begin{figure}[!htb] 
\setlength{\abovecaptionskip}{0cm}
\setlength{\belowcaptionskip}{0cm}
\centering 
\includegraphics[width=\textwidth]{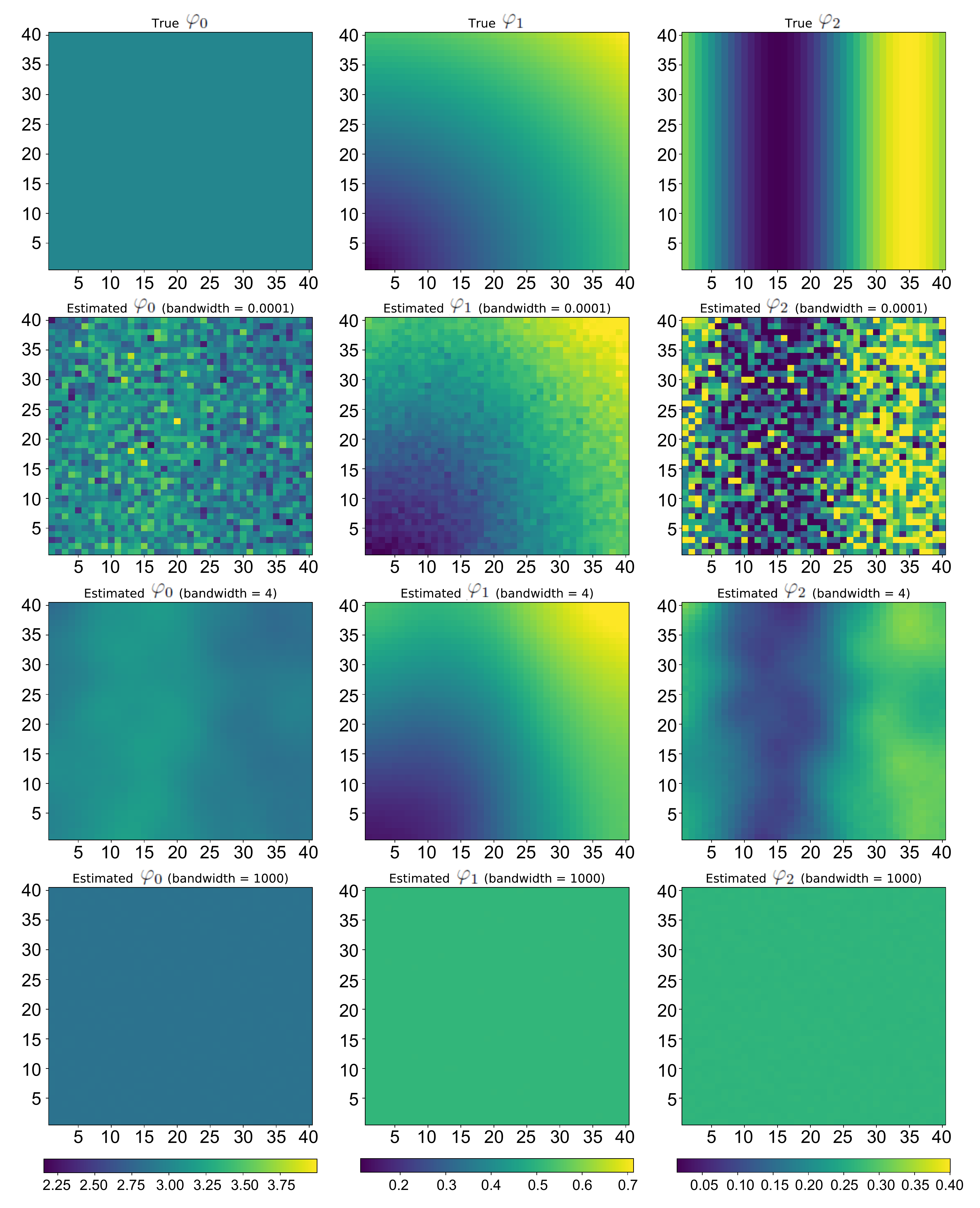} 
\caption[Heatmap of the true values and estimated means for the coefficients $\varphi_0$, $\varphi_1$ and $\varphi_2$ by the Bayesian GWR model, with geographic bandwidth $\eta = 0.0001,1,$ and $20$.]{\textbf{Heatmap of the true values and estimated means for the coefficients $\varphi_0$, $\varphi_1$ and $\varphi_2$ by the Bayesian GWR model, with geographic bandwidth $\eta = 0.0001,4$ and $1000$.}} 
\label{F5}
\end{figure}

We then ran the model on the complete dataset without excluding any observations. For each location, we ran 10 chains independently for $4\times10^3$ iterations, discarding the first $1\times10^3$ samples as burn-in, so that the change of the value of the estimated elpd was smaller than 0.05 (trace plot in supplementary materials). \if 0 The upper and lower bounds of trace plots reveal that the empirical SMI distribution tends to have lower variance with higher geographical bandwidth. \fi The true values and estimated means for coefficients $(\varphi_0,\varphi_1,\varphi_2)$, when $\eta = 0.0001,4,$ and $1000$, are shown in Figure \ref{F5}. When $\eta=0.0001$, estimation at each location relies almost exclusively on data from that location, so the estimated coefficients vary considerably across spatial locations: the connection between locations is almost completely ``cut''. Furthermore, some estimates are extreme because excluding neighbouring samples means only a small number of samples are used by the model. These results reveal the nature of using a small bandwidth in a GWR model, as has also been discussed previously \citep{doi:10.1139/X08-091}. In contrast, when $\eta=1000$, we can see the estimated coefficients are almost constant across geographic locations, due to the large bandwidth that assumes samples from neighbouring locations are very similar to samples from the location of interest. Finally, the estimates using the optimal bandwidth $\eta=4$ are close to the true values across all geographic locations. 

\begin{figure}[t] 
\setlength{\abovecaptionskip}{0cm}
\setlength{\belowcaptionskip}{0cm}
\centering 
\includegraphics[width=\textwidth]{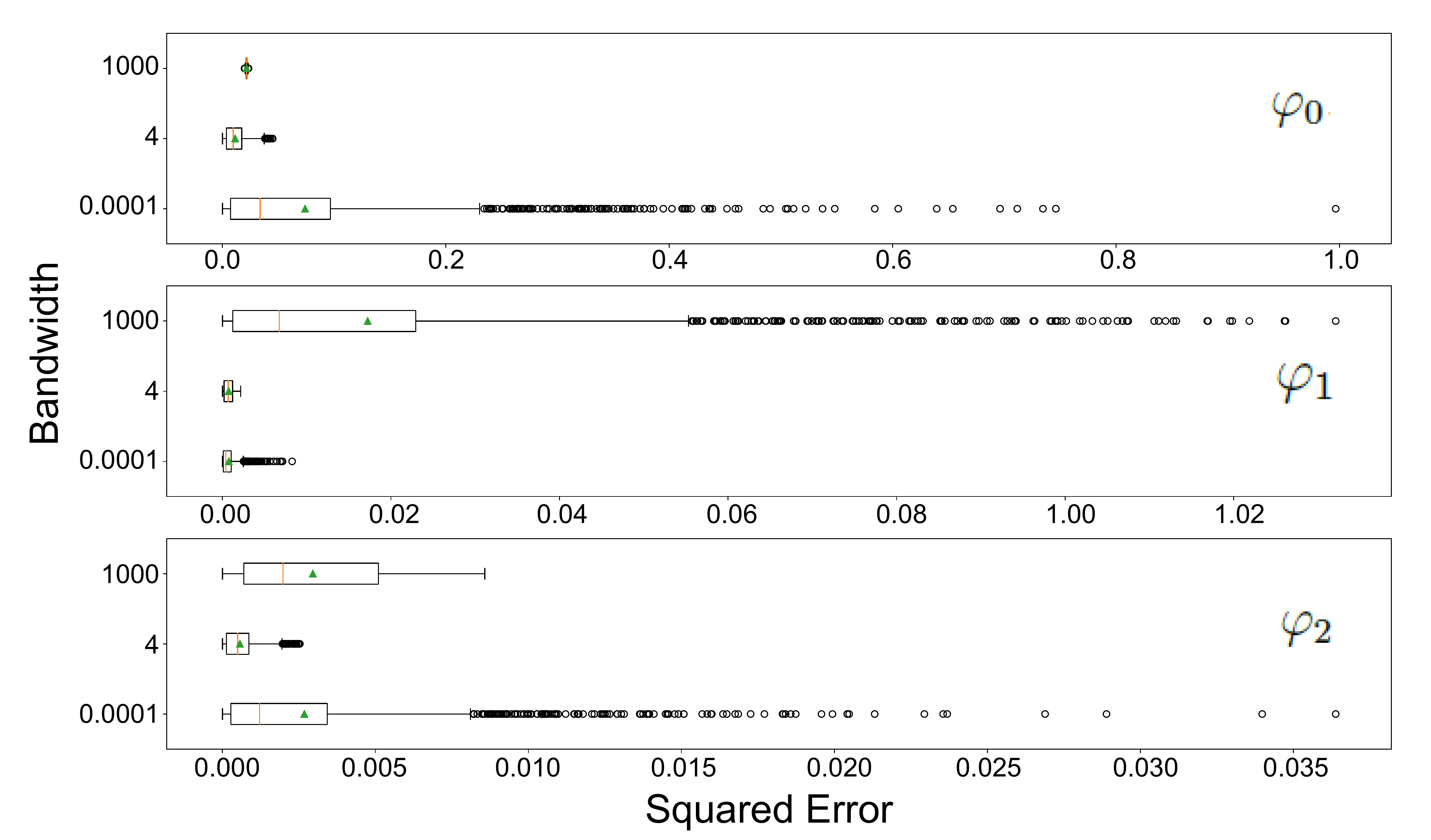} 
\caption[Boxplots of the squared error of the estimated mean coefficients $\varphi_0$, $\varphi_1$ and $\varphi_2$ under the Bayesian GWR model across geographic locations, with geographic bandwidth $\eta = 0.0001,4,$ and $1000$.]{\textbf{Boxplots of the squared error of the estimated mean coefficients $\varphi_0$, $\varphi_1$ and $\varphi_2$ under the Bayesian GWR model across geographic locations, with geographic bandwidth $\eta = 0.0001,4,$ and $1000$.} The orange line and green triangle indicate the median and mean squared error.} 
\label{F6}
\end{figure}

Figure \ref{F6} shows boxplots of the squared error between the Bayesian GWR estimated means and the true values of the three coefficients across all geographic locations. The true $\varphi_0$ is constant, therefore a large bandwidth that incorporates more samples will have lower mean squared error. Hence, the model with $\eta=4$ provides good estimation of $\varphi_0$. In contrast, the model with $\eta=0.0001$ fails to estimate the true value of $\varphi_0$ because the sample size at each location is not sufficient to enable precise estimation. Moreover, the model with $\eta=1000$ has a significant bias because it incorporates too much information from other locations which have considerably different data generating processes to the location of interest. For $\varphi_1$ and $\varphi_2$ which do vary geographically, the model with $\eta=1000$ as expected performs poorly because the model assumes little geographic variation. The model with $\eta=0.0001$ also performs poorly due to the insufficient sample size at each individual location. Overall, the model with the optimal bandwidth $\eta=4$ performs the best in mean squared error. The supplementary material contains further discussion of the estimation error.

\section{Application to Real Data}
It has been shown in epidemiological studies that there is a global variation in the seasonal activity of the influenza virus \citep[e.g.,][]{finkelman2007global,10.1093/infdis/jis467,lam2019comparative}. In particular, there are normally clear and consistent influenza epidemic peaks during the winter in the high-latitude regions \citep{cox2000global}, whereas seasonal transmission patterns are unclear in low-latitude (subtropical/tropical) regions \citep{viboud2006influenza,li2019global}. This suggests that transmission and viability of the influenza virus is linked with atmospheric conditions: the regular occurrence of influenza epidemic in temperate regions is largely attributed to the exposure of cold and dry environments \citep[e.g.,][]{lowen2007influenza,Lowen7692,deyle2016global,chong2020association}. However, this relationship is weaker in subtropical/tropical regions \citep{tamerius2013environmental}. In this section, we apply the Bayesian GWR model to a human influenza dataset to assess spatial variation in the association between the occurrence of influenza and two major climatic factors (temperature and precipitation).

We used monthly, country-level human influenza surveillance data between January 2010 and December 2014 from the World Health Organization FluNet (\url{https://www.who.int/tools/flunet}). We selected 20 countries of similar size and with relatively comprehensive influenza records. We selected 16 European countries to represent the temperate region (Austria; Belgium; Bosnia and Herzegovina; Croatia; Czech; France; Germany; Hungary; Italy; Luxembourg; Netherlands; Poland; Romania; Slovakia; Slovenia; UK) and 4 South-East Asian countries to represent the tropical region (Cambodia; Laos; Thailand; Vietnam). We used the geographical center coordinates $(u_i,v_i)$, $i=1,\cdots,20$ of each country as the geographical coordinates. The dataset contains the number of positive cases $Y_{i,t}$ and total number of tests $N_{i,t}$ in country $i=1,\cdots,20$ during month $t$. 
The temperature $X_{i1,t}$ (degrees Celsius) and amount of precipitation $X_{i2,t}$ (mm/month) during month $t$ in country $i$ were obtained from CRUCY \citep{harris2014updated}.

\begin{figure}[t] 
\setlength{\abovecaptionskip}{0cm}
\setlength{\belowcaptionskip}{0cm}
\centering 
\includegraphics[width=\textwidth]{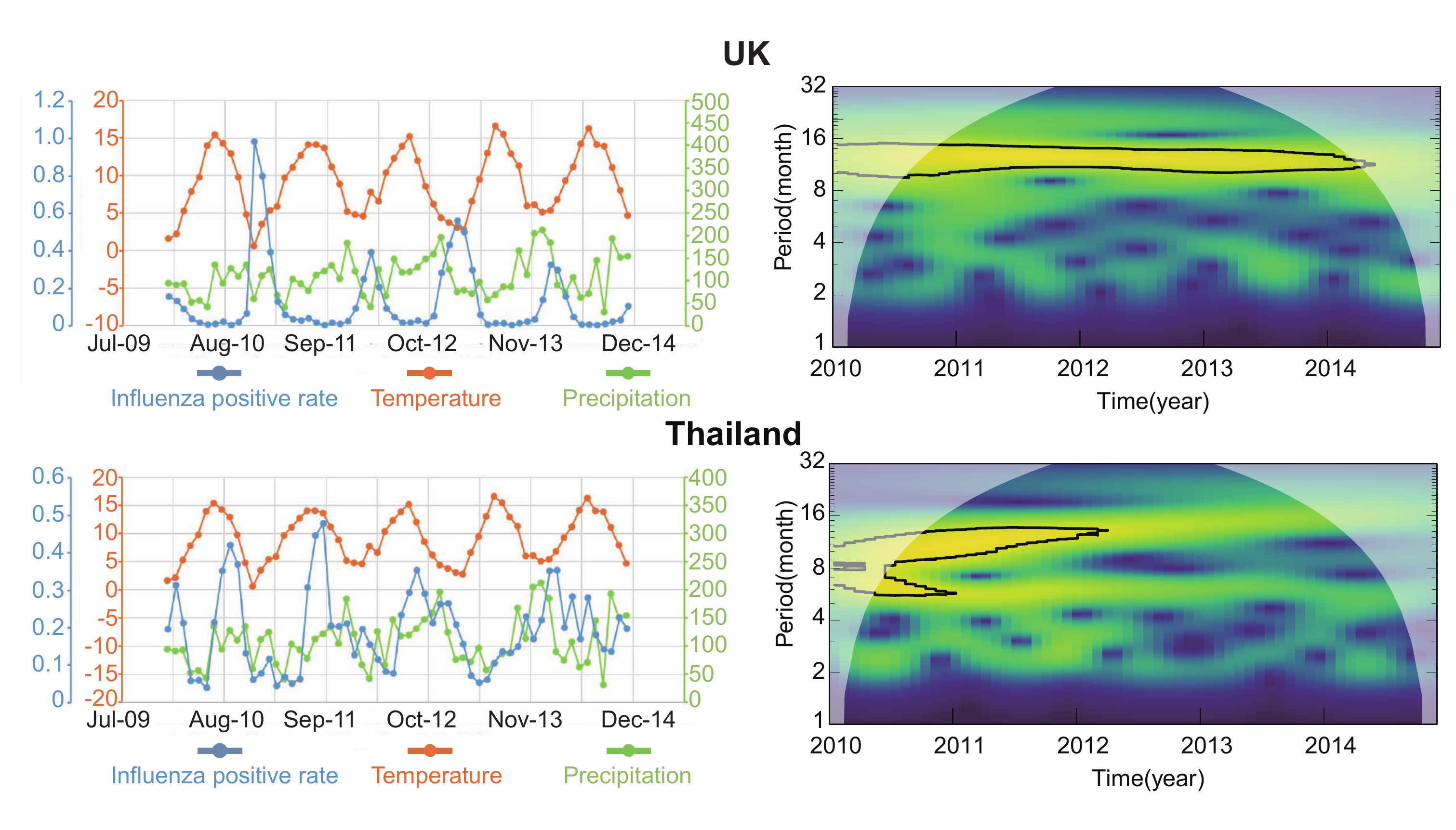} 
\caption[Association between influenza, temperature and precipitation in the UK and Thailand.]{\textbf{Association between influenza, temperature and precipitation in the UK and Thailand.} Left panel: Monthly influenza positivity rate, temperature (degrees Celsius) and precipitation (mm/month). Right panel: Wavelet power spectrum (absolute square of the wavelet transform) of the influenza activity. The black line surrounds the significant area (p-value $<$ 0.01), where the power spectrum is significantly large than the power spectrum of random noise.} 
\label{F8}
\end{figure}

The countries we included show distinct patterns of influenza activities. Figure \ref{F8} shows, for the UK and Thailand, the monthly influenza positivity rate, temperature, precipitation and the corresponding wavelet analysis of the periodicity of influenza activity. In the UK, we can observe that the peak of influenza activity is consistent with the winter season in the UK and a clear negative correlation can be observed between influenza positivity rate and temperature. The relationship visually appears less strong for precipitation. In contrast, in Thailand influenza has a more variable peak time and the relationship with temperature and precipitation is not clear. To further quantify the distinct seasonality of influenza activities between two countries for better understanding of the underlying geographical difference, we conducted a separate (exploratory) wavelet analysis using WaveletComp in R. This decomposes the influenza time series into numerous wavelets, each with a distinct frequency. The degree to which influenza follows a particular periodicity can be assessed by the magnitude of the corresponding wavelet. This reveals clear evidence of periodicity of between 10-15 months in all years in the UK, whereas there is no consistent periodicity in Thailand (Figure \ref{F8}). This highlights the potential geographical variation of the influenza activities, suggesting a GWR model is appropriate.

\begin{figure}[!htb] 
\setlength{\abovecaptionskip}{0cm}
\setlength{\belowcaptionskip}{0cm}
\centering 
\includegraphics[width=\textwidth]{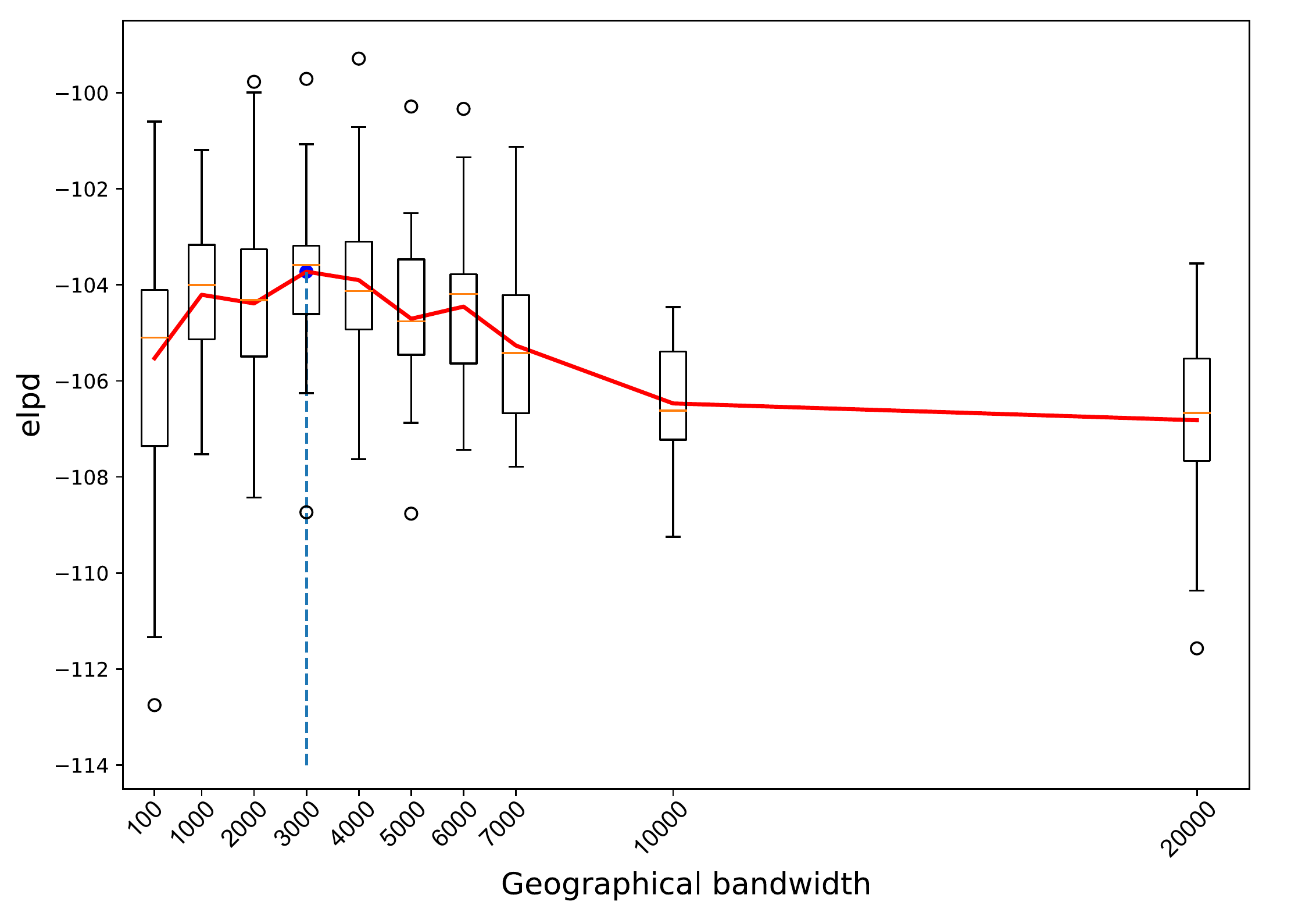} 
\caption[ELPD against geographical bandwidth.]{\textbf{ELPD against geographical bandwidth.} Each boxplot represents the elpd estimates from 30 chains across 20 countries. The red line is the average elpd estimates across the 30 chains. The blue dashed line indicates the optimal bandwidth.} 
\label{F9}
\end{figure}

In our Bayesian GWR model, we assumed that the number of positive cases $Y_{i,t}$ follows a negative binomial distribution, as in \eqref{E2}, except that the total number of tests $N_{i,t}$ was embedded into the link function and spherical distance was calculated using the haversine formula. The mean and variance of $Y_{i,t}$ are:
\[
\resizebox{\textwidth}{!}{$
\begin{aligned}
\mathbb{E}(Y_{i,t}|X_{i1,t},X_{i2,t}) &= \exp(\log(N_{i,t})+ \varphi_0(u_i,v_i)+\varphi_1(u_i,v_i)X_{i1,t}+\varphi_2(u_i,v_i)X_{i2,t}), \\
\textbf{Var}(Y_{i,t}|X_{i1,t},X_{i2,t}) &= \mathbb{E}(Y_{i,t}|X_{i1,t},X_{i2,t}) + \theta_i\mathbb{E}^2(Y_{i,t}|X_{i1,t},X_{i2,t}).
\end{aligned}
$}
\]
We considered each of the 20 countries separately, with each of the following geographical bandwidths $\eta =  100$, 1000, 2000, 3000, 4000, 5000, 6000, 7000, 10000 and 20000 (kilometres) for a Gaussian kernel. These choices of bandwidth cover a broad range of different assumptions regarding the impact of neighbouring countries. For each country, we randomly left-out 50\% of the observations to use as a test set. We ran 30 independent MCMC chains, and after discarding the first $3\times 10^3$ samples, we drew $10^4$ samples from the Bayesian GWR posterior. Figure \ref{F9} shows the estimated elpd for each bandwidth across the whole space, suggesting that the optimal choice of the bandwidth from the candidate set is 3000. This bandwidth indicates that there is spatial variation of the underlying association across the countries we selected. Note that, the range of 3000 kilometres has roughly spans either Europe or South-East Asia but not both, meaning that spatial non-stationarity was detected between these two regions but the spatial non-stationarity is not significant within the two regions. 

\begin{figure}[t] 
\setlength{\abovecaptionskip}{0cm}
\setlength{\belowcaptionskip}{0cm}
\centering 
\includegraphics[width=\textwidth]{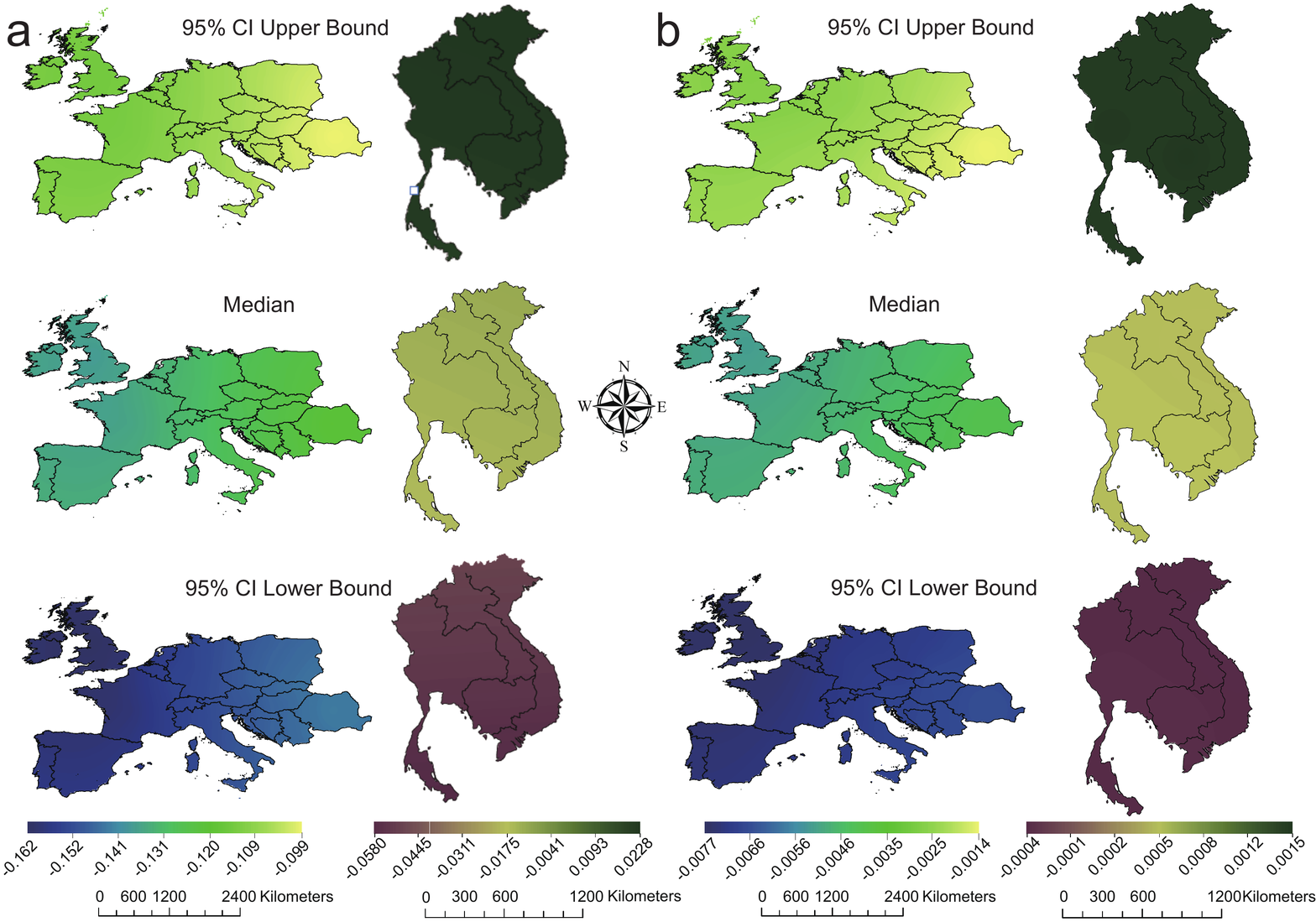} 
\caption[Geographical variation of the estimated $\varphi$ for temperature (a) and precipitation (b).]{\textbf{Median (middle panel), lower (bottom panel) and upper (top panel) 95\% credible intervals for $\varphi$ for European (left column) and East Asian (right column) countries. Panel (a) shows estimates for $\varphi_1$ for temperature (a); panel (b) shows estimates for $\varphi_2$ for precipitation.} The color and map scales are listed at the bottom. } 
\label{F10}
\end{figure}

We applied the model in all 20 countries independently, using the whole dataset and with bandwidth $\eta=3000$. We ran 20 independent MCMC chains for each country, and retained $10^4$ samples after discarding the first $3\times 10^3$ samples as burn-in. The pooled samples drawn from the Bayesian GWR posterior for $\varphi$ for temperature and precipitation were used to estimate the median, lower and upper $95\%$ bound of credible interval (CI) for each country. Figure \ref{F10} shows the results, after applying kriging interpolation with ArcGIS Version 10.7. These estimates imply that in European countries a negative association exists between influenza and both temperature and precipitation. That is, influenza transmission tends to be more prevalent during the cold and dry season. In contrast, there is no significant association in the south-east Asian countries. These conclusions are consistent with previous findings \citep[e.g.,][]{tamerius2013environmental}.

\section{Conclusions}
We have introduced and extended the SMI model and the candidate distribution selection technique to the field of geographic information science (GIS). Currently, a Bayesian approach for GWR models is only available for the Gaussian linear regression \citep{SUBEDI20182574011,doi:10.1177/0160017620959823}. We therefore elucidate the theoretical validity of applying a Bayesian approach to generalized GWR models and reveal the essential link between the Bayesian GWR model and cutting or manipulating feedback. The motivation of Bayesian GWR model is to decrease the random error at the expense of introducing systematic error. This is realized by incorporating observations from neighbouring locations. The geographically weighted kernel manipulates the information provided by extra observations. The optimal geographical bandwidth $\eta$ balances the trade-off between two types of error. Our model can also be applied for the Gaussian distribution with $\theta$ being the standard deviation. We note that our Bayesian GWR for Gaussian is different to the Bayesian GWR proposed by \cite{doi:10.1177/0160017620959823}. This is because our model is based on the weighted log-likelihood while \cite{doi:10.1177/0160017620959823} is based on a weighted least squares approach. Specifically for the Gaussian distribution, these two models may be equivalent if the parameter of interest is only $\varphi$ because only the exponential term of the likelihood, which is proportional to the residual sum of squares when log-likelihood is used, contains $\varphi$. However, they are different if $\theta$ is also considered.

GWR models in a frequentist framework require tedious mathematical derivation of the estimator to obtain estimates of the uncertainty of the parameter estimates, which may not be always accessible. In contrast, the Bayesian GWR model provides easily obtainable and straightforward measures of the uncertainty of the parameter estimates given the posterior samples. Furthermore, the Bayesian nature of this model means that prior knowledge can be easily introduced into the model. Unlike the SVC generalized linear model, which may require Monte Carlo sampling in a high-dimensional parameter space, the Bayesian GWR model requires only sampling separately for each location, meaning the dimension of the parameter does not scale with the number of locations, regardless of the generalized linear model used. Regarding computation, unlike the SVC model and other standard Bayesian spatial methods which require sequential sampling of parameters for all locations, the Bayesian GWR model can easily benefit from the availability of parallelization due to the separate inference for each location.

While most GWR models have considered spatially smooth parameters (coefficients), the GWR literature has not previously consider the more general case when some of the parameters are locally unique but not spatially smooth e.g. linear regression, negative binomial regression or beta regression involving a parameter akin to $\theta_i$ in \eqref{E2}. Hence, our model can be viewed as an extension of the conventional GWR models that is able to simultaneously deal with (1) spatially smooth and (2) locally unique but not spatially smooth parameters.

The SMI model was previously only established for the two module case, i.e. with a single cut. In this study, we extend it to a special case of multiple cuts when information from suspect modules are manipulated via a deterministic functional form controlled by a single kernel bandwidth.

Several limitations of the current model are left for future investigation. First, the current model selects the optimal bandwidth using cross-validation. This can be computationally expensive since it requires multiple partitions of the set of observations $Y_{i,1:m}$ for each location $i$. Second, although the current model can infer the parameter $\theta$, this inference may suffer from insufficient observations because the inference of $\theta$ only depends on observations from the location of interest as shown in \eqref{E13}. Third, our model uses a globally fixed geographical bandwidth. This could be problematic when the true data generating process varies considerably within some areas but only varies to a small degree within other areas; or when some elements of the regression coefficient $\varphi$ have a large geographical variation whereas other elements of $\varphi$ have a small geographical variation. Spatially-varying bandwidth or parameter-specific distance metrics have been proposed for standard GWR models \citep{leong2017modification,doi:10.1080/24694452.2017.1352480,doi:10.1080/13658816.2016.1263731,math9182343}, but the extension of these methods within a Bayesian framework is not straightforward computationally because a basic implementation would involve repeated evaluation of the geographically weighted kernel for all locations.

\section*{Supplementary Materials}
The supplementary appendix contains all technical proofs of results stated in the paper.

\section*{Acknowledgement}
Yang Liu was supported by a Cambridge International Scholarship from the Cambridge Commonwealth, European and International Trust. Robert J.B. Goudie was funded by the UK Medical Research Council [programme code MC\textunderscore UU\textunderscore 00002/2].

%\bibliography{references}  %%% Remove comment to use the external .bib file (using bibtex).
%%% and comment out the ``thebibliography'' section.

%%% Comment out this section when you \bibliography{references} is enabled.
%\bibliographystyle{apalike}
%\bibliography{reference}

\begin{thebibliography}{88}
\newcommand{\enquote}[1]{``#1''}
\expandafter\ifx\csname natexlab\endcsname\relax\def\natexlab#1{#1}\fi
\expandafter\ifx\csname url\endcsname\relax
  \def\url#1{{\tt #1}}\fi
\expandafter\ifx\csname urlprefix\endcsname\relax\def\urlprefix{URL }\fi
\ifx\endbibitem\undefined \let\endbibitem\relax\fi

\bibitem[{Afroughi et~al.(2011)Afroughi, Faghihzadeh, Khaledi, Motlagh, and
  Hajizadeh}]{doi:10.1080/02664763.2011.570315}
Afroughi, S., Faghihzadeh, S., Khaledi, M.~J., Motlagh, M.~G., and Hajizadeh,
  E. (2011).
\newblock \enquote{{Analysis of clustered spatially correlated binary data
  using autologistic model and Bayesian method with an application to dental
  caries of 3–5-year-old children}.}
\newblock {\em Journal of Applied Statistics\/}, 38(12): 2763--2774.
\newline\urlprefix\url{https://doi.org/10.1080/02664763.2011.570315}
\endbibitem

\bibitem[{Alquier et~al.(2016)Alquier, Ridgway, and Chopin}]{JMLR:v17:15-290}
Alquier, P., Ridgway, J., and Chopin, N. (2016).
\newblock \enquote{{On the properties of variational approximations of Gibbs
  posteriors}.}
\newblock {\em Journal of Machine Learning Research\/}, 17(236): 1--41.
\newline\urlprefix\url{http://jmlr.org/papers/v17/15-290.html}
\endbibitem

\bibitem[{Arendt et~al.(2012)Arendt, Apley, and
  Chen}]{arendt2012quantification}
Arendt, P.~D., Apley, D.~W., and Chen, W. (2012).
\newblock \enquote{{Quantification of model uncertainty: Calibration, model
  discrepancy, and identifiability}.}
\newblock {\em Journal of Mechanical Design\/}, 134(10).
\newblock 100908.
\newline\urlprefix\url{https://doi.org/10.1115/1.4007390}
\endbibitem

\bibitem[{Azziz~Baumgartner et~al.(2012)Azziz~Baumgartner, Dao, Nasreen,
  Bhuiyan, Mah-E-Muneer, Mamun, Sharker, Zaman, Cheng, Klimov, Widdowson,
  Uyeki, Luby, Mounts, and Bresee}]{10.1093/infdis/jis467}
Azziz~Baumgartner, E., Dao, C.~N., Nasreen, S., Bhuiyan, M.~U., Mah-E-Muneer,
  S., Mamun, A.~A., Sharker, M. A.~Y., Zaman, R.~U., Cheng, P.-Y., Klimov,
  A.~I., Widdowson, M.-A., Uyeki, T.~M., Luby, S.~P., Mounts, A., and Bresee,
  J. (2012).
\newblock \enquote{{Seasonality, timing, and climate drivers of influenza
  activity worldwide}.}
\newblock {\em The Journal of Infectious Diseases\/}, 206(6): 838--846.
\newline\urlprefix\url{https://doi.org/10.1093/infdis/jis467}
\endbibitem

\bibitem[{Banerjee et~al.(2008)Banerjee, Gelfand, Finley, and
  Sang}]{https://doi.org/10.1111/j.1467-9868.2008.00663.x}
Banerjee, S., Gelfand, A.~E., Finley, A.~O., and Sang, H. (2008).
\newblock \enquote{Gaussian predictive process models for large spatial data
  sets.}
\newblock {\em Journal of the Royal Statistical Society: Series B (Statistical
  Methodology)\/}, 70(4): 825--848.
\newline\urlprefix\url{https://rss.onlinelibrary.wiley.com/doi/abs/10.1111/j.1467-9868.2008.00663.x}
\endbibitem

\bibitem[{Berrocal et~al.(2010)Berrocal, Gelfand, and
  Holland}]{berrocal2010spatio}
Berrocal, V.~J., Gelfand, A.~E., and Holland, D.~M. (2010).
\newblock \enquote{A spatio-temporal downscaler for output from numerical
  models.}
\newblock {\em Journal of Agricultural, Biological, and Environmental
  Statistics\/}, 15(2): 176--197.
\endbibitem

\bibitem[{Bhattacharya et~al.(2019)Bhattacharya, Pati, and
  Yang}]{bhattacharya2019}
Bhattacharya, A., Pati, D., and Yang, Y. (2019).
\newblock \enquote{{Bayesian fractional posteriors}.}
\newblock {\em The Annals of Statistics\/}, 47(1): 39 -- 66.
\newline\urlprefix\url{https://doi.org/10.1214/18-AOS1712}
\endbibitem

\bibitem[{Bissiri et~al.(2016)Bissiri, Holmes, and
  Walker}]{https://doi.org/10.1111/rssb.12158}
Bissiri, P.~G., Holmes, C.~C., and Walker, S.~G. (2016).
\newblock \enquote{A general framework for updating belief distributions.}
\newblock {\em Journal of the Royal Statistical Society: Series B (Statistical
  Methodology)\/}, 78(5): 1103--1130.
\newline\urlprefix\url{https://rss.onlinelibrary.wiley.com/doi/abs/10.1111/rssb.12158}
\endbibitem

\bibitem[{Biswas et~al.(2015)Biswas, Roy, Majumder, and
  Basu}]{https://doi.org/10.1002/sta4.80}
Biswas, A., Roy, T., Majumder, S., and Basu, A. (2015).
\newblock \enquote{A new weighted likelihood approach.}
\newblock {\em Stat\/}, 4(1): 97--107.
\newline\urlprefix\url{https://onlinelibrary.wiley.com/doi/abs/10.1002/sta4.80}
\endbibitem

\bibitem[{Blangiardo et~al.(2011)Blangiardo, Hansell, and
  Richardson}]{BLANGIARDO2011379}
Blangiardo, M., Hansell, A., and Richardson, S. (2011).
\newblock \enquote{A Bayesian model of time activity data to investigate health
  effect of air pollution in time series studies.}
\newblock {\em Atmospheric Environment\/}, 45(2): 379 -- 386.
\newline\urlprefix\url{http://www.sciencedirect.com/science/article/pii/S1352231010008642}
\endbibitem

\bibitem[{Brunsdon et~al.(1996)Brunsdon, Fotheringham, and
  Charlton}]{doi:10.1111/j.1538-4632.1996.tb00936.x}
Brunsdon, C., Fotheringham, A.~S., and Charlton, M.~E. (1996).
\newblock \enquote{Geographically weighted regression: A method for exploring
  spatial nonstationarity.}
\newblock {\em Geographical Analysis\/}, 28(4): 281--298.
\newline\urlprefix\url{https://onlinelibrary.wiley.com/doi/abs/10.1111/j.1538-4632.1996.tb00936.x}
\endbibitem

\bibitem[{Cai et~al.(2000)Cai, Fan, and
  Li}]{doi:10.1080/01621459.2000.10474280}
Cai, Z., Fan, J., and Li, R. (2000).
\newblock \enquote{Efficient estimation and inferences for varying-coefficient
  models.}
\newblock {\em Journal of the American Statistical Association\/}, 95(451):
  888--902.
\newline\urlprefix\url{https://www.tandfonline.com/doi/abs/10.1080/01621459.2000.10474280}
\endbibitem

\bibitem[{Carmona and Nicholls(2020)}]{carmona2020semi}
Carmona, C. and Nicholls, G. (2020).
\newblock \enquote{{Semi-modular inference: Enhanced learning in multi-modular
  models by tempering the influence of components}.}
\newblock In Chiappa, S. and Calandra, R. (eds.), {\em Proceedings of the
  Twenty Third International Conference on Artificial Intelligence and
  Statistics\/}, volume 108 of {\em Proceedings of Machine Learning
  Research\/}, 4226--4235. PMLR.
\endbibitem

\bibitem[{Chen et~al.(2012)Chen, Deng, Yang, and
  Matthews}]{https://doi.org/10.1111/j.1538-4632.2012.00841.x}
Chen, V. Y.-J., Deng, W.-S., Yang, T.-C., and Matthews, S.~A. (2012).
\newblock \enquote{{Geographically weighted quantile regression (GWQR): An
  application to U.S. mortality data}.}
\newblock {\em Geographical Analysis\/}, 44(2): 134--150.
\newline\urlprefix\url{https://onlinelibrary.wiley.com/doi/abs/10.1111/j.1538-4632.2012.00841.x}
\endbibitem

\bibitem[{Chernozhukov and Hong(2003)}]{CHERNOZHUKOV2003293}
Chernozhukov, V. and Hong, H. (2003).
\newblock \enquote{An MCMC approach to classical estimation.}
\newblock {\em Journal of Econometrics\/}, 115(2): 293 -- 346.
\newline\urlprefix\url{http://www.sciencedirect.com/science/article/pii/S0304407603001003}
\endbibitem

\bibitem[{Chong et~al.(2020)Chong, Lee, Bialasiewicz, Chen, Smith, Choy,
  Krajden, Jalal, Jennings, Alexander et~al.}]{chong2020association}
Chong, K.~C., Lee, T.~C., Bialasiewicz, S., Chen, J., Smith, D.~W., Choy,
  W.~S., Krajden, M., Jalal, H., Jennings, L., Alexander, B., et~al. (2020).
\newblock \enquote{{Association between meteorological variations and
  activities of influenza A and B across different climate zones: A
  multi-region modelling analysis across the globe}.}
\newblock {\em Journal of Infection\/}, 80(1): 84--98.
\endbibitem

\bibitem[{Cox and Subbarao(2000)}]{cox2000global}
Cox, N.~J. and Subbarao, K. (2000).
\newblock \enquote{Global epidemiology of influenza: Past and present.}
\newblock {\em Annual Review of Medicine\/}, 51(1): 407--421.
\newline\urlprefix\url{https://doi.org/10.1146/annurev.med.51.1.407}
\endbibitem

\bibitem[{{da Silva} and {de Oliveira Lima}(2017)}]{DASILVA2017279}
{da Silva}, A.~R. and {de Oliveira Lima}, A. (2017).
\newblock \enquote{Geographically weighted beta regression.}
\newblock {\em Spatial Statistics\/}, 21: 279 -- 303.
\newline\urlprefix\url{http://www.sciencedirect.com/science/article/pii/S2211675317300179}
\endbibitem

\bibitem[{da~Silva and Rodrigues(2014)}]{da2014geographically}
da~Silva, A.~R. and Rodrigues, T. C.~V. (2014).
\newblock \enquote{Geographically weighted negative binomial
  regression—incorporating overdispersion.}
\newblock {\em Statistics and Computing\/}, 24(5): 769--783.
\endbibitem

\bibitem[{Dalalyan and Tsybakov(2008)}]{dalalyan2008aggregation}
Dalalyan, A. and Tsybakov, A.~B. (2008).
\newblock \enquote{Aggregation by exponential weighting, sharp PAC-Bayesian
  bounds and sparsity.}
\newblock {\em Machine Learning\/}, 72(1-2): 39--61.
\endbibitem

\bibitem[{Deyle et~al.(2016)Deyle, Maher, Hernandez, Basu, and
  Sugihara}]{deyle2016global}
Deyle, E.~R., Maher, M.~C., Hernandez, R.~D., Basu, S., and Sugihara, G.
  (2016).
\newblock \enquote{Global environmental drivers of influenza.}
\newblock {\em Proceedings of the National Academy of Sciences\/}, 113(46):
  13081--13086.
\newline\urlprefix\url{https://www.pnas.org/content/113/46/13081}
\endbibitem

\bibitem[{Duan and Li(2016)}]{7517288}
Duan, S.-B. and Li, Z.-L. (2016).
\newblock \enquote{Spatial Downscaling of MODIS Land Surface Temperatures Using
  Geographically Weighted Regression: Case Study in Northern China.}
\newblock {\em IEEE Transactions on Geoscience and Remote Sensing\/}, 54(11):
  6458--6469.
\endbibitem

\bibitem[{Dunson and Taylor(2005)}]{doi:10.1080/10485250500039049}
Dunson, D.~B. and Taylor, J.~A. (2005).
\newblock \enquote{Approximate Bayesian inference for quantiles.}
\newblock {\em Journal of Nonparametric Statistics\/}, 17(3): 385--400.
\newline\urlprefix\url{https://doi.org/10.1080/10485250500039049}
\endbibitem

\bibitem[{Finkelman et~al.(2007)Finkelman, Viboud, Koelle, Ferrari, Bharti, and
  Grenfell}]{finkelman2007global}
Finkelman, B.~S., Viboud, C., Koelle, K., Ferrari, M.~J., Bharti, N., and
  Grenfell, B.~T. (2007).
\newblock \enquote{{Global patterns in seasonal activity of influenza A/H3N2,
  A/H1N1, and B from 1997 to 2005: Viral coexistence and latitudinal
  gradients}.}
\newblock {\em PLOS ONE\/}, 2(12): 1--10.
\newline\urlprefix\url{https://doi.org/10.1371/journal.pone.0001296}
\endbibitem

\bibitem[{Finley et~al.(2007)Finley, Banerjee, and Carlin}]{finley2007spbayes}
Finley, A.~O., Banerjee, S., and Carlin, B.~P. (2007).
\newblock \enquote{spBayes: an R package for univariate and multivariate
  hierarchical point-referenced spatial models.}
\newblock {\em Journal of Statistical Software\/}, 19(4): 1.
\endbibitem

\bibitem[{Fotheringham et~al.(1996)Fotheringham, Charlton, and
  Brunsdon}]{doi:10.1080/02693799608902100}
Fotheringham, A.~S., Charlton, M., and Brunsdon, C. (1996).
\newblock \enquote{The geography of parameter space: an investigation of
  spatial non-stationarity.}
\newblock {\em International Journal of Geographical Information Systems\/},
  10(5): 605--627.
\newline\urlprefix\url{https://doi.org/10.1080/02693799608902100}
\endbibitem

\bibitem[{Fotheringham et~al.(2017)Fotheringham, Yang, and
  Kang}]{doi:10.1080/24694452.2017.1352480}
Fotheringham, A.~S., Yang, W., and Kang, W. (2017).
\newblock \enquote{Multiscale geographically weighted regression (MGWR).}
\newblock {\em Annals of the American Association of Geographers\/}, 107(6):
  1247--1265.
\newline\urlprefix\url{https://doi.org/10.1080/24694452.2017.1352480}
\endbibitem

\bibitem[{Frank et~al.(2019)Frank, Massman, Ewers, and
  Williams}]{https://doi.org/10.1029/2018WR023054}
Frank, J.~M., Massman, W.~J., Ewers, B.~E., and Williams, D.~G. (2019).
\newblock \enquote{Bayesian analyses of 17 winters of water vapor fluxes show
  bark beetles reduce sublimation.}
\newblock {\em Water Resources Research\/}, 55(2): 1598--1623.
\newline\urlprefix\url{https://agupubs.onlinelibrary.wiley.com/doi/abs/10.1029/2018WR023054}
\endbibitem

\bibitem[{Friel and
  Pettitt(2008)}]{https://doi.org/10.1111/j.1467-9868.2007.00650.x}
Friel, N. and Pettitt, A.~N. (2008).
\newblock \enquote{Marginal likelihood estimation via power posteriors.}
\newblock {\em Journal of the Royal Statistical Society: Series B (Statistical
  Methodology)\/}, 70(3): 589--607.
\newline\urlprefix\url{https://rss.onlinelibrary.wiley.com/doi/abs/10.1111/j.1467-9868.2007.00650.x}
\endbibitem

\bibitem[{Fuglstad et~al.(2015)Fuglstad, Lindgren, Simpson, and
  Rue}]{10.2307/24311007}
Fuglstad, G.-A., Lindgren, F., Simpson, D., and Rue, H. (2015).
\newblock \enquote{{Exploring a new class of non-stationary spatial Gaussian
  random fields with varying local anisotropy}.}
\newblock {\em Statistica Sinica\/}, 25(1): 115--133.
\newline\urlprefix\url{http://www.jstor.org/stable/24311007}
\endbibitem

\bibitem[{Gelfand and
  Banerjee(2017)}]{doi:10.1146/annurev-statistics-060116-054155}
Gelfand, A.~E. and Banerjee, S. (2017).
\newblock \enquote{Bayesian modeling and analysis of geostatistical data.}
\newblock {\em Annual Review of Statistics and Its Application\/}, 4(1):
  245--266.
\newline\urlprefix\url{https://doi.org/10.1146/annurev-statistics-060116-054155}
\endbibitem

\bibitem[{Gelfand et~al.(2003)Gelfand, Kim, Sirmans, and
  Banerjee}]{doi:10.1198/016214503000170}
Gelfand, A.~E., Kim, H.-J., Sirmans, C.~F., and Banerjee, S. (2003).
\newblock \enquote{Spatial modeling with spatially varying coefficient
  processes.}
\newblock {\em Journal of the American Statistical Association\/}, 98(462):
  387--396.
\newline\urlprefix\url{https://doi.org/10.1198/016214503000170}
\endbibitem

\bibitem[{Gelman et~al.(2014)Gelman, Hwang, and
  Vehtari}]{gelman2014understanding}
Gelman, A., Hwang, J., and Vehtari, A. (2014).
\newblock \enquote{{Understanding predictive information criteria for Bayesian
  models}.}
\newblock {\em Statistics and Computing\/}, 24(6): 997--1016.
\endbibitem

\bibitem[{{Geman} and {Geman}(1984)}]{4767596}
{Geman}, S. and {Geman}, D. (1984).
\newblock \enquote{{Stochastic relaxation, Gibbs distributions, and the
  Bayesian restoration of images}.}
\newblock {\em IEEE Transactions on Pattern Analysis and Machine
  Intelligence\/}, PAMI-6(6): 721--741.
\endbibitem

\bibitem[{Grünwald and van Ommen(2017)}]{grunwald2017}
Grünwald, P. and van Ommen, T. (2017).
\newblock \enquote{{Inconsistency of Bayesian inference for misspecified linear
  models, and a proposal for repairing it}.}
\newblock {\em Bayesian Analysis\/}, 12(4): 1069 -- 1103.
\newline\urlprefix\url{https://doi.org/10.1214/17-BA1085}
\endbibitem

\bibitem[{Guo et~al.(2008)Guo, Ma, and Zhang}]{doi:10.1139/X08-091}
Guo, L., Ma, Z., and Zhang, L. (2008).
\newblock \enquote{Comparison of bandwidth selection in application of
  geographically weighted regression: a case study.}
\newblock {\em Canadian Journal of Forest Research\/}, 38(9): 2526--2534.
\newline\urlprefix\url{https://doi.org/10.1139/X08-091}
\endbibitem

\bibitem[{Harris et~al.(2014)Harris, Jones, Osborn, and
  Lister}]{harris2014updated}
Harris, I., Jones, P., Osborn, T., and Lister, D. (2014).
\newblock \enquote{Updated high-resolution grids of monthly climatic
  observations – the CRU TS3.10 Dataset.}
\newblock {\em International Journal of Climatology\/}, 34(3): 623--642.
\newline\urlprefix\url{https://rmets.onlinelibrary.wiley.com/doi/abs/10.1002/joc.3711}
\endbibitem

\bibitem[{Holmes and Walker(2017)}]{10.1093/biomet/asx010}
Holmes, C.~C. and Walker, S.~G. (2017).
\newblock \enquote{{Assigning a value to a power likelihood in a general
  Bayesian model}.}
\newblock {\em Biometrika\/}, 104(2): 497--503.
\newline\urlprefix\url{https://doi.org/10.1093/biomet/asx010}
\endbibitem

\bibitem[{Hu and Zidek(2002)}]{https://doi.org/10.2307/3316141}
Hu, F. and Zidek, J.~V. (2002).
\newblock \enquote{The weighted likelihood.}
\newblock {\em Canadian Journal of Statistics\/}, 30(3): 347--371.
\newline\urlprefix\url{https://onlinelibrary.wiley.com/doi/abs/10.2307/3316141}
\endbibitem

\bibitem[{Hu et~al.(2021)Hu, Lu, Zhang, Jiang, and Shi}]{math9182343}
Hu, X., Lu, Y., Zhang, H., Jiang, H., and Shi, Q. (2021).
\newblock \enquote{Selection of the bandwidth matrix in spatial varying
  coefficient models to detect anisotropic regression relationships.}
\newblock {\em Mathematics\/}, 9(18).
\newline\urlprefix\url{https://www.mdpi.com/2227-7390/9/18/2343}
\endbibitem

\bibitem[{Jacob et~al.(2017)Jacob, Murray, Holmes, and
  Robert}]{jacob2017better}
Jacob, P.~E., Murray, L.~M., Holmes, C.~C., and Robert, C.~P. (2017).
\newblock \enquote{Better together? Statistical learning in models made of
  modules.}
\newblock {\em arXiv preprint arXiv:1708.08719\/}.
\endbibitem

\bibitem[{Jiang and Tanner(2008)}]{jiang2008}
Jiang, W. and Tanner, M.~A. (2008).
\newblock \enquote{{Gibbs posterior for variable selection in high-dimensional
  classification and data mining}.}
\newblock {\em The Annals of Statistics\/}, 36(5): 2207 -- 2231.
\newline\urlprefix\url{https://doi.org/10.1214/07-AOS547}
\endbibitem

\bibitem[{Kaplan and Chen(2012)}]{kaplan2012two}
Kaplan, D. and Chen, J. (2012).
\newblock \enquote{A two-step Bayesian approach for propensity score analysis:
  Simulations and case study.}
\newblock {\em Psychometrika\/}, 77(3): 581--609.
\endbibitem

\bibitem[{Lam et~al.(2019)Lam, Tang, Lai, Zaraket, Dbaibo, Bialasiewicz, Tozer,
  Heraud, Drews, Hachette et~al.}]{lam2019comparative}
Lam, T.~T., Tang, J.~W., Lai, F.~Y., Zaraket, H., Dbaibo, G., Bialasiewicz, S.,
  Tozer, S., Heraud, J.-M., Drews, S.~J., Hachette, T., et~al. (2019).
\newblock \enquote{Comparative global epidemiology of influenza, respiratory
  syncytial and parainfluenza viruses, 2010--2015.}
\newblock {\em Journal of Infection\/}, 79(4): 373--382.
\endbibitem

\bibitem[{Leong and Yue(2017)}]{leong2017modification}
Leong, Y.-Y. and Yue, J.~C. (2017).
\newblock \enquote{A modification to geographically weighted regression.}
\newblock {\em International Journal of Health Geographics\/}, 16(1): 11.
\endbibitem

\bibitem[{Lever et~al.(2013)Lever, Laviolette, and Shawe-Taylor}]{LEVER20134}
Lever, G., Laviolette, F., and Shawe-Taylor, J. (2013).
\newblock \enquote{Tighter PAC-Bayes bounds through distribution-dependent
  priors.}
\newblock {\em Theoretical Computer Science\/}, 473: 4 -- 28.
\newblock Special Issue on Algorithmic Learning Theory.
\newline\urlprefix\url{http://www.sciencedirect.com/science/article/pii/S0304397512009346}
\endbibitem

\bibitem[{Li and Sang(2019)}]{doi:10.1080/01621459.2018.1529595}
Li, F. and Sang, H. (2019).
\newblock \enquote{Spatial Homogeneity Pursuit of Regression Coefficients for
  Large Datasets.}
\newblock {\em Journal of the American Statistical Association\/}, 114(527):
  1050--1062.
\newline\urlprefix\url{https://doi.org/10.1080/01621459.2018.1529595}
\endbibitem

\bibitem[{Li et~al.(2019)Li, Reeves, Wang, Bassat, Brooks, Cohen, Moore, Nunes,
  Rath, Campbell et~al.}]{li2019global}
Li, Y., Reeves, R.~M., Wang, X., Bassat, Q., Brooks, W.~A., Cohen, C., Moore,
  D.~P., Nunes, M., Rath, B., Campbell, H., et~al. (2019).
\newblock \enquote{Global patterns in monthly activity of influenza virus,
  respiratory syncytial virus, parainfluenza virus, and metapneumovirus: a
  systematic analysis.}
\newblock {\em The Lancet Global Health\/}, 7(8): e1031--e1045.
\endbibitem

\bibitem[{Li and Fotheringham(2020)}]{doi:10.1080/13658816.2020.1720692}
Li, Z. and Fotheringham, A.~S. (2020).
\newblock \enquote{Computational improvements to multi-scale geographically
  weighted regression.}
\newblock {\em International Journal of Geographical Information Science\/},
  34(7): 1378--1397.
\newline\urlprefix\url{https://doi.org/10.1080/13658816.2020.1720692}
\endbibitem

\bibitem[{Lin(2010)}]{lin2010estimating}
Lin, P.-S. (2010).
\newblock \enquote{Estimating equations for separable spatial-temporal binary
  data.}
\newblock {\em Environmental and Ecological Statistics\/}, 17(4): 543--557.
\endbibitem

\bibitem[{Liu et~al.(2009)Liu, Bayarri, Berger et~al.}]{liu2009modularization}
Liu, F., Bayarri, M., Berger, J., et~al. (2009).
\newblock \enquote{{Modularization in Bayesian analysis, with emphasis on
  analysis of computer models}.}
\newblock {\em Bayesian Analysis\/}, 4(1): 119--150.
\endbibitem

\bibitem[{Liu and Goudie(2022)}]{liu2020stochastic}
Liu, Y. and Goudie, R. J.~B. (2022).
\newblock \enquote{Stochastic approximation cut algorithm for inference in
  modularized Bayesian models.}
\newblock {\em Statistics and Computing\/}, 32(1): 1--15.
\endbibitem

\bibitem[{Liu et~al.(2018)Liu, Lam, Wu, and Lam}]{liu2018geographically}
Liu, Y., Lam, K.-F., Wu, J.~T., and Lam, T. T.-Y. (2018).
\newblock \enquote{Geographically weighted temporally correlated logistic
  regression model.}
\newblock {\em Scientific Reports\/}, 8(1): 1--14.
\endbibitem

\bibitem[{Lowen et~al.(2007)Lowen, Mubareka, Steel, and
  Palese}]{lowen2007influenza}
Lowen, A.~C., Mubareka, S., Steel, J., and Palese, P. (2007).
\newblock \enquote{Influenza virus transmission is dependent on relative
  humidity and temperature.}
\newblock {\em PLOS Pathogens\/}, 3(10): 1--7.
\newline\urlprefix\url{https://doi.org/10.1371/journal.ppat.0030151}
\endbibitem

\bibitem[{Lowen and Steel(2014)}]{Lowen7692}
Lowen, A.~C. and Steel, J. (2014).
\newblock \enquote{Roles of humidity and temperature in shaping influenza
  seasonality.}
\newblock {\em Journal of Virology\/}, 88(14): 7692--7695.
\newline\urlprefix\url{https://jvi.asm.org/content/88/14/7692}
\endbibitem

\bibitem[{Lu et~al.(2017)Lu, Brunsdon, Charlton, and
  Harris}]{doi:10.1080/13658816.2016.1263731}
Lu, B., Brunsdon, C., Charlton, M., and Harris, P. (2017).
\newblock \enquote{Geographically weighted regression with parameter-specific
  distance metrics.}
\newblock {\em International Journal of Geographical Information Science\/},
  31(5): 982--998.
\newline\urlprefix\url{https://doi.org/10.1080/13658816.2016.1263731}
\endbibitem

\bibitem[{Lunn et~al.(2009)Lunn, Best, Spiegelhalter, Graham, and
  Neuenschwander}]{lunn2009combining}
Lunn, D., Best, N., Spiegelhalter, D., Graham, G., and Neuenschwander, B.
  (2009).
\newblock \enquote{Combining MCMC with ‘sequential’ PKPD modelling.}
\newblock {\em Journal of Pharmacokinetics and Pharmacodynamics\/}, 36(1):
  19—38.
\newline\urlprefix\url{https://doi.org/10.1007/s10928-008-9109-1}
\endbibitem

\bibitem[{Ma et~al.(2020)Ma, Xue, and Hu}]{doi:10.1177/0160017620959823}
Ma, Z., Xue, Y., and Hu, G. (2020).
\newblock \enquote{{Geographically weighted regression analysis for spatial
  economics data: A Bayesian recourse}.}
\newblock {\em International Regional Science Review\/}, 44(5): 582--604.
\newline\urlprefix\url{https://doi.org/10.1177/0160017620959823}
\endbibitem

\bibitem[{Markatou(2000)}]{https://doi.org/10.1111/j.0006-341X.2000.00483.x}
Markatou, M. (2000).
\newblock \enquote{Mixture models, robustness, and the weighted likelihood
  methodology.}
\newblock {\em Biometrics\/}, 56(2): 483--486.
\newline\urlprefix\url{https://onlinelibrary.wiley.com/doi/abs/10.1111/j.0006-341X.2000.00483.x}
\endbibitem

\bibitem[{Marques et~al.(2020)Marques, Klein, and Kneib}]{MARQUES2020100386}
Marques, I., Klein, N., and Kneib, T. (2020).
\newblock \enquote{Non-stationary spatial regression for modelling monthly
  precipitation in Germany.}
\newblock {\em Spatial Statistics\/}, 40: 100386.
\newline\urlprefix\url{http://www.sciencedirect.com/science/article/pii/S221167531930137X}
\endbibitem

\bibitem[{Martin et~al.(2017)Martin, Mess, and Walker}]{martin2017}
Martin, R., Mess, R., and Walker, S.~G. (2017).
\newblock \enquote{Empirical Bayes posterior concentration in sparse
  high-dimensional linear models.}
\newblock {\em Bernoulli\/}, 23(3): 1822 -- 1847.
\newline\urlprefix\url{https://doi.org/10.3150/15-BEJ797}
\endbibitem

\bibitem[{Mayfield et~al.(2018)Mayfield, Lowry, Watson, Kama, Nilles, and
  Lau}]{MAYFIELD2018e223}
Mayfield, H.~J., Lowry, J.~H., Watson, C.~H., Kama, M., Nilles, E.~J., and Lau,
  C.~L. (2018).
\newblock \enquote{Use of geographically weighted logistic regression to
  quantify spatial variation in the environmental and sociodemographic drivers
  of leptospirosis in Fiji: a modelling study.}
\newblock {\em The Lancet Planetary Health\/}, 2(5): e223--e232.
\newline\urlprefix\url{https://www.sciencedirect.com/science/article/pii/S2542519618300664}
\endbibitem

\bibitem[{McCandless et~al.(2010)McCandless, Douglas, Evans, and
  Smeeth}]{McCandlessDouglasEvansSmeeth2010}
McCandless, L.~C., Douglas, I.~J., Evans, S.~J., and Smeeth, L. (2010).
\newblock \enquote{{Cutting feedback in Bayesian regression adjustment for the
  propensity score}.}
\newblock {\em The International Journal of Biostatistics\/}, 6(2): 16.
\newline\urlprefix\url{https://doi.org/10.2202/1557-4679.1205}
\endbibitem

\bibitem[{Miller and Dunson(2019)}]{doi:10.1080/01621459.2018.1469995}
Miller, J.~W. and Dunson, D.~B. (2019).
\newblock \enquote{{Robust Bayesian inference via coarsening}.}
\newblock {\em Journal of the American Statistical Association\/}, 114(527):
  1113--1125.
\newline\urlprefix\url{https://doi.org/10.1080/01621459.2018.1469995}
\endbibitem

\bibitem[{Mohammed et~al.(2022)Mohammed, Ravikumar, Warner, Patel, Bakas, Rao,
  and Jain}]{Mohammed33}
Mohammed, S., Ravikumar, V., Warner, E., Patel, S., Bakas, S., Rao, A., and
  Jain, R. (2022).
\newblock \enquote{Quantifying T2-FLAIR Mismatch Using Geographically Weighted
  Regression and Predicting Molecular Status in Lower-Grade Gliomas.}
\newblock {\em American Journal of Neuroradiology\/}, 43(1): 33--39.
\newline\urlprefix\url{http://www.ajnr.org/content/43/1/33}
\endbibitem

\bibitem[{Mu et~al.(2018)Mu, Wang, and Wang}]{https://doi.org/10.1002/env.2485}
Mu, J., Wang, G., and Wang, L. (2018).
\newblock \enquote{Estimation and inference in spatially varying coefficient
  models.}
\newblock {\em Environmetrics\/}, 29(1): e2485.
\newline\urlprefix\url{https://onlinelibrary.wiley.com/doi/abs/10.1002/env.2485}
\endbibitem

\bibitem[{Nakaya et~al.(2005)Nakaya, Fotheringham, Brunsdon, and
  Charlton}]{doi:10.1002/sim.2129}
Nakaya, T., Fotheringham, A.~S., Brunsdon, C., and Charlton, M. (2005).
\newblock \enquote{Geographically weighted Poisson regression for disease
  association mapping.}
\newblock {\em Statistics in Medicine\/}, 24(17): 2695--2717.
\newline\urlprefix\url{https://onlinelibrary.wiley.com/doi/abs/10.1002/sim.2129}
\endbibitem

\bibitem[{Paez et~al.(2005)Paez, Gamerman, and
  De~Oliveira}]{paez2005interpolation}
Paez, M.~S., Gamerman, D., and De~Oliveira, V. (2005).
\newblock \enquote{Interpolation performance of a spatio-temporal model with
  spatially varying coefficients: application to PM 10 concentrations in Rio de
  Janeiro.}
\newblock {\em Environmental and Ecological Statistics\/}, 12(2): 169--193.
\endbibitem

\bibitem[{Plummer(2015)}]{plummer2015cuts}
Plummer, M. (2015).
\newblock \enquote{{Cuts in Bayesian graphical models}.}
\newblock {\em Statistics and Computing\/}, 25(1): 37--43.
\endbibitem

\bibitem[{Reich et~al.(2010)Reich, Fuentes, Herring, and
  Evenson}]{https://doi.org/10.1111/j.1541-0420.2009.01333.x}
Reich, B.~J., Fuentes, M., Herring, A.~H., and Evenson, K.~R. (2010).
\newblock \enquote{Bayesian Variable Selection for Multivariate Spatially
  Varying Coefficient Regression.}
\newblock {\em Biometrics\/}, 66(3): 772--782.
\newline\urlprefix\url{https://onlinelibrary.wiley.com/doi/abs/10.1111/j.1541-0420.2009.01333.x}
\endbibitem

\bibitem[{Rubin(2008)}]{rubin2008}
Rubin, D.~B. (2008).
\newblock \enquote{{For objective causal inference, design trumps analysis}.}
\newblock {\em The Annals of Applied Statistics\/}, 2(3): 808 -- 840.
\newline\urlprefix\url{https://doi.org/10.1214/08-AOAS187}
\endbibitem

\bibitem[{Subedi et~al.(2018)Subedi, Zhang, and Zhen}]{SUBEDI20182574011}
Subedi, N., Zhang, L., and Zhen, Z. (2018).
\newblock \enquote{{Bayesian geographically weighted regression and its
  application for local modeling of relationships between tree variables}.}
\newblock {\em iForest - Biogeosciences and Forestry\/}, (5): 542--552.
\newline\urlprefix\url{https://iforest.sisef.org/contents/?id=ifor2574-011}
\endbibitem

\bibitem[{Sugasawa and Murakami(2021)}]{SUGASAWA2021100525}
Sugasawa, S. and Murakami, D. (2021).
\newblock \enquote{Spatially clustered regression.}
\newblock {\em Spatial Statistics\/}, 44: 100525.
\newline\urlprefix\url{https://www.sciencedirect.com/science/article/pii/S221167532100035X}
\endbibitem

\bibitem[{Tamerius et~al.(2013)Tamerius, Shaman, Alonso, Bloom-Feshbach, Uejio,
  Comrie, and Viboud}]{tamerius2013environmental}
Tamerius, J.~D., Shaman, J., Alonso, W.~J., Bloom-Feshbach, K., Uejio, C.~K.,
  Comrie, A., and Viboud, C. (2013).
\newblock \enquote{Environmental predictors of seasonal influenza epidemics
  across temperate and tropical climates.}
\newblock {\em PLOS Pathogens\/}, 9(3): 1--12.
\newline\urlprefix\url{https://doi.org/10.1371/journal.ppat.1003194}
\endbibitem

\bibitem[{Tasyurek and Celik(2020)}]{TASYUREK2020258}
Tasyurek, M. and Celik, M. (2020).
\newblock \enquote{RNN-GWR: A geographically weighted regression approach for
  frequently updated data.}
\newblock {\em Neurocomputing\/}, 399: 258--270.
\newline\urlprefix\url{https://www.sciencedirect.com/science/article/pii/S0925231220302484}
\endbibitem

\bibitem[{Tobler(1970)}]{tobler1970computer}
Tobler, W.~R. (1970).
\newblock \enquote{A computer movie simulating urban growth in the Detroit
  region.}
\newblock {\em Economic Geography\/}, 46(sup1): 234--240.
\endbibitem

\bibitem[{Utazi et~al.(2019)Utazi, Thorley, Alegana, Ferrari, Nilsen,
  Takahashi, Metcalf, Lessler, and Tatem}]{doi:10.1177/0962280218797362}
Utazi, C., Thorley, J., Alegana, V., Ferrari, M., Nilsen, K., Takahashi, S.,
  Metcalf, C., Lessler, J., and Tatem, A. (2019).
\newblock \enquote{A spatial regression model for the disaggregation of areal
  unit based data to high-resolution grids with application to vaccination
  coverage mapping.}
\newblock {\em Statistical Methods in Medical Research\/}, 28(10-11):
  3226--3241.
\newline\urlprefix\url{https://doi.org/10.1177/0962280218797362}
\endbibitem

\bibitem[{Viboud et~al.(2006)Viboud, Alonso, and
  Simonsen}]{viboud2006influenza}
Viboud, C., Alonso, W.~J., and Simonsen, L. (2006).
\newblock \enquote{Influenza in tropical regions.}
\newblock {\em PLOS Medicine\/}, 3(4): e89.
\newline\urlprefix\url{https://doi.org/10.1371/journal.pmed.0030089}
\endbibitem

\bibitem[{Walker and Hjort(2001)}]{https://doi.org/10.1111/1467-9868.00314}
Walker, S. and Hjort, N.~L. (2001).
\newblock \enquote{{On Bayesian consistency}.}
\newblock {\em Journal of the Royal Statistical Society: Series B (Statistical
  Methodology)\/}, 63(4): 811--821.
\newline\urlprefix\url{https://rss.onlinelibrary.wiley.com/doi/abs/10.1111/1467-9868.00314}
\endbibitem

\bibitem[{Wang et~al.(2019)Wang, Shi, Fang, and Feng}]{WANG201995}
Wang, S., Shi, C., Fang, C., and Feng, K. (2019).
\newblock \enquote{Examining the spatial variations of determinants of
  energy-related CO2 emissions in China at the city level using Geographically
  Weighted Regression Model.}
\newblock {\em Applied Energy\/}, 235: 95--105.
\newline\urlprefix\url{https://www.sciencedirect.com/science/article/pii/S0306261918316520}
\endbibitem

\bibitem[{Windle et~al.(2009)Windle, Rose, Devillers, and
  Fortin}]{10.1093/icesjms/fsp224}
Windle, M. J.~S., Rose, G.~A., Devillers, R., and Fortin, M.-J. (2009).
\newblock \enquote{{Exploring spatial non-stationarity of fisheries survey data
  using geographically weighted regression (GWR): an example from the Northwest
  Atlantic}.}
\newblock {\em ICES Journal of Marine Science\/}, 67(1): 145--154.
\newline\urlprefix\url{https://doi.org/10.1093/icesjms/fsp224}
\endbibitem

\bibitem[{Wu(2020)}]{WU2020121089}
Wu, D. (2020).
\newblock \enquote{Spatially and temporally varying relationships between
  ecological footprint and influencing factors in China’s provinces Using
  Geographically Weighted Regression (GWR).}
\newblock {\em Journal of Cleaner Production\/}, 261: 121089.
\newline\urlprefix\url{https://www.sciencedirect.com/science/article/pii/S0959652620311367}
\endbibitem

\bibitem[{Wu et~al.(2021)Wu, Wang, Du, Huang, Zhang, and
  Liu}]{doi:10.1080/13658816.2020.1775836}
Wu, S., Wang, Z., Du, Z., Huang, B., Zhang, F., and Liu, R. (2021).
\newblock \enquote{Geographically and temporally neural network weighted
  regression for modeling spatiotemporal non-stationary relationships.}
\newblock {\em International Journal of Geographical Information Science\/},
  35(3): 582--608.
\newline\urlprefix\url{https://doi.org/10.1080/13658816.2020.1775836}
\endbibitem

\bibitem[{Zellner(1988)}]{doi:10.1080/00031305.1988.10475585}
Zellner, A. (1988).
\newblock \enquote{{Optimal information processing and Bayes's theorem}.}
\newblock {\em The American Statistician\/}, 42(4): 278--280.
\newline\urlprefix\url{https://www.tandfonline.com/doi/abs/10.1080/00031305.1988.10475585}
\endbibitem

\bibitem[{Zhang(2006)}]{1614067}
Zhang, T. (2006).
\newblock \enquote{Information-theoretic upper and lower bounds for statistical
  estimation.}
\newblock {\em IEEE Transactions on Information Theory\/}, 52(4): 1307--1321.
\endbibitem

\bibitem[{Zhu et~al.(2005)Zhu, Huang, and Wu}]{zhu2005modeling}
Zhu, J., Huang, H.-C., and Wu, J. (2005).
\newblock \enquote{{Modeling spatial-temporal binary data using Markov random
  fields}.}
\newblock {\em Journal of Agricultural, Biological, and Environmental
  Statistics\/}, 10(2): 212--225.
\newline\urlprefix\url{http://www.jstor.org/stable/27595556}
\endbibitem

\bibitem[{Zigler and Dominici(2014)}]{doi:10.1080/01621459.2013.869498}
Zigler, C.~M. and Dominici, F. (2014).
\newblock \enquote{Uncertainty in propensity score estimation: Bayesian methods
  for variable selection and model-averaged causal effects.}
\newblock {\em Journal of the American Statistical Association\/}, 109(505):
  95--107.
\newline\urlprefix\url{https://doi.org/10.1080/01621459.2013.869498}
\endbibitem

\bibitem[{Zigler et~al.(2013)Zigler, Watts, Yeh, Wang, Coull, and
  Dominici}]{zigler2013model}
Zigler, C.~M., Watts, K., Yeh, R.~W., Wang, Y., Coull, B.~A., and Dominici, F.
  (2013).
\newblock \enquote{{Model feedback in Bayesian propensity score estimation}.}
\newblock {\em Biometrics\/}, 69(1): 263--273.
\newline\urlprefix\url{https://onlinelibrary.wiley.com/doi/abs/10.1111/j.1541-0420.2012.01830.x}
\endbibitem

\end{thebibliography}

\appendix
\appendixpage
\addappheadtotoc

\renewcommand{\thefigure}{A\arabic{figure}}

\setcounter{figure}{0}

\renewcommand{\thetable}{A\arabic{table}}

\setcounter{table}{0}

\newpage

\section{Estimation error of Bayesian GWR model}
To further explore the estimation error of the Bayesian GWR model, given the simulation in the main text (Section 5), we look at the estimation of individual location $\hat{\varphi}(u,v)$ to the true value $\varphi(u,v)$. An ideal model should give estimation which achieves $\hat{\varphi}=\varphi$. However, estimation will inevitably affected by error. Notice that, although we have run a Bayesian GWR model independently at each of the 1600 locations, they share the same geographical bandwidth $\eta$. Therefore, it is reasonable to assume that the degree of error introduced due to the model should be similar for all locations. Here, we assume the following equation:
\[
\hat{\varphi}(u,v) = F(\varphi(u,v))+\varepsilon(u,v),
\]
where $F$ is a unknown deterministic function and $\varepsilon(u,v)$ $i.i.d$ follows an arbitrary distribution with mean 0 and variance $\sigma^2$. The term $\varepsilon(u,v)$ describes the random error that naturally arises from samples due to the randomness of observations and can be reduced by increasing the sample size. The squared error is $(F(\varphi(u,v))+\varepsilon(u,v)-\varphi(u,v))^2$. The function $F$ describes the systematic error that is due to the misspecification (i.e., the use of samples from neighbouring locations). When a coefficient varies geographically, this systematic error can not be removed if we include neighbouring samples. If there is no systematic error (i.e., we do not borrow any sample from neighbouring locations or coefficient does not vary across the space), then $F(\varphi(u,v))=\varphi(u,v)$.

\begin{figure}[h] 
\setlength{\abovecaptionskip}{0cm}
\setlength{\belowcaptionskip}{0cm}
\includegraphics[width=\textwidth]{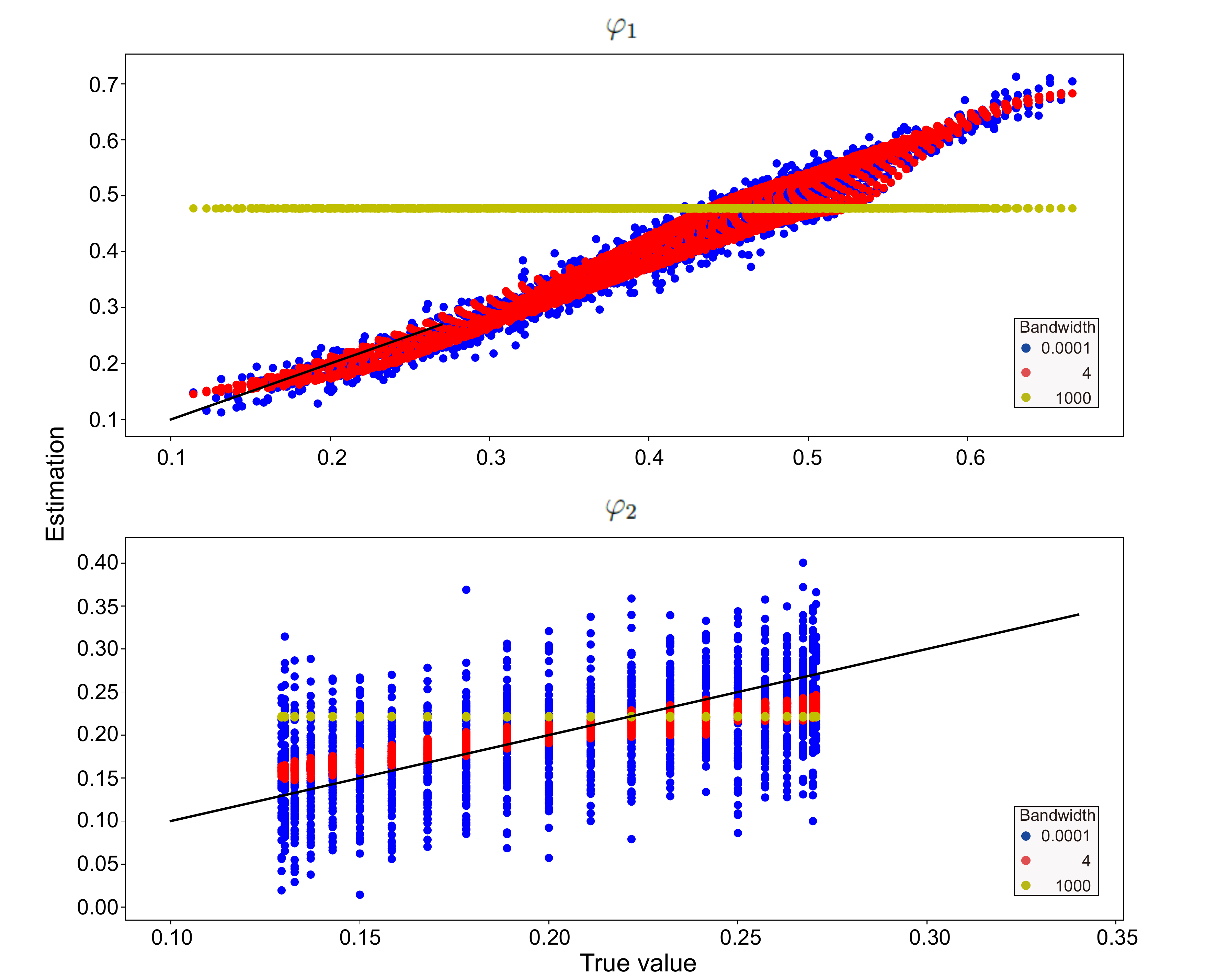} 
\caption[Scatter plot of the estimated mean coefficient for $\psi_1$ and $\psi_2$ under the Bayesian GWR model against the true value at each of the 1600 geographic locations.]{\textbf{Scatter plot of the estimated mean coefficient for $\psi_1$ and $\psi_2$ under the Bayesian GWR model against the true value at each of the 1600 geographic locations.} Results are shown for geographic bandwidths $\eta=0.0001$ (blue), $\eta=4$ (red) and $\eta=1000$ (yellow). The diagonal benchmark line $\hat{\varphi}=\varphi$ indicates where the estimates should be centred around if there is no systematic error.} 
\label{F7}
\end{figure}

Figure \ref{F7} shows a scatter plot of the estimated mean coefficients at each of the geographic locations for each bandwidth choice, against their true values. The estimates when $\eta = 0.0001$ distribute evenly around the true values because this model emphasizes local characteristics, but with large variance due to the large random error due to the insufficient number of samples used by this model. In contrast, the estimates when $\eta = 1000$ are relatively horizontal because this model assumes coefficients are relatively constant across the geographical space, leading to a large deviation from the true values due to the systematic error caused by including too much information from  neighbouring geographic locations. The model with the optimal bandwidth $\eta=4$ has less  systematic error than $\eta=1000$, and much smaller random error than $\eta = 0.0001$.

Given the clear linear trend for all bandwidth choices, we assume a linear form $F(\varphi(u,v))=a+b\varphi(u,v)$ for the systematic error, and summarise the results via the linear regression coefficients (Table \ref{T1}). We first consider the systematic error. When $\eta=0.0001$ and $\eta=4$, the intercept is close to 0 and the slope is close to 1, indicating the systematic error is very small. In contrast when $\eta=1000$, the intercept differs from 0 and the slope is clearly not close to 1. This confirms the systematic error we discussed before. Specifically, the slope goes to 0 when $\eta=1000$. This again reveals the fact that larger geographical bandwidth ignores geographical variation. Now we consider the random error. It is clear that, for both $\varphi_1$ and $\varphi_2$, the model with $\eta=1000$ gives the smallest random error $\sigma$ and $\sigma$ increases as $\eta$ decreases. This trend reveals the varying pattern of the dispersion thanks to the changing of the sample size. In summary, the model with $\eta=4$ balances both systematic and random error.

\begin{table}
\centering
\begin{tabular}{c c c c c}
\toprule
Coefficient & Bandwidth & $\hat{a}$ & $\hat{b}$ & $\hat{\sigma}$ \\
\hline
 & 0.0001 & -0.0428 & 1.1041 & 0.0249 \\

$\varphi_1$ & 4 & -0.0445  & 1.1169 & 0.0236 \\

 & 1000 & 0.4775  & 0.0003 & 0.0003 \\

 & 0.0001 & 0.0597 & 0.6857 & 0.0492 \\

$\varphi_2$ & 4 & 0.0901  & 0.5395 & 0.0061 \\

 & 1000 & $0.2213 $ & 0.0001 & 0.0003 \\
\toprule
\end{tabular}
\caption{Estimated intercept $\hat{a}$, slope $\hat{b}$ and standard deviation $\hat{\sigma}$ under the linear model of the estimated mean coefficients for each choice of geographic bandwidth.}
\label{T1}
\end{table}

\section{Proofs of the Main Text}
\subsection{Proof of the Lemma 1}
\begin{proof}
We prove the lemma when $n=2$; the proof can be easily extended to case when $n>2$ by induction.

Given the posterior $p(\theta_{0:n},\varphi_0|Y_{0:n,1:m})$, we have
\[
p(\theta_0,\theta_1,\theta_2,\varphi_0|Y_{0:2,1:m})=p(\theta_2|\theta_0,\theta_1,\varphi_0,Y_{0:2,1:m})p(\theta_0,\theta_1,\varphi_0|Y_{0:2,1:m}).
\]
Then by conditional independence of $\theta_2$ and $(Y_{0:1,1:m},\theta_{0:1})$ given $\varphi_0$, we have
\[
p(\theta_0,\theta_1,\theta_2,\varphi_0|Y_{0:2,1:m})=p(\theta_2|Y_{2,1:m},\varphi_0)p(\theta_0,\theta_1,\varphi_0|Y_{0:2,1:m}).
\]
For the term $p(\theta_0,\theta_1,\varphi_0|Y_{0:2,1:m})$, we have
\[
\begin{aligned}
p(\theta_0,\theta_1,\varphi_0|Y_{0:2,1:m})&=p(\theta_1|\theta_0,\varphi_0,Y_{0:2,1:m})p(\theta_0,\varphi_0|Y_{0:2,1:m}) \\
&=\int p(\theta_1|\theta_0,\theta_2,\varphi_0,Y_{0:2,1:m})p(\theta_2|\theta_0,\varphi_0,Y_{0:2,1:m})d\theta_2 \ p(\theta_0,\varphi_0|Y_{0:2,1:m}).
\end{aligned}
\]
Similarly, by conditional independence of $\theta_1$ and $(Y_{-1,1:m},\theta_{-1})$ given $\varphi$
\[
\begin{aligned}
p(\theta_0,\theta_1,\varphi_0|Y_{0:2,1:m})&=\int p(\theta_1|Y_{1,1:m},\varphi_0)p(\theta_2|\theta_0,\varphi_0,Y_{0:2,1:m})d\theta_2 \ p(\theta_0,\varphi_0|Y_{0:2,1:m}) \\
&=p(\theta_1|Y_{1,1:m},\varphi_0)p(\theta_0,\varphi_0|Y_{0:2,1:m}).
\end{aligned}
\]
Hence, we have:
\[
p(\theta_0,\theta_1,\theta_2,\varphi_0|Y_{0:2,1:m})= p(\theta_2|Y_{2,1:m},\varphi_0)p(\theta_1|Y_{1,1:m},\varphi_0) p(\theta_0,\varphi_0|Y_{0:2,1:m}).
\]

\end{proof}

\subsection{Proof of the Theorem 1}
\begin{proof}
We first notice that the following equality holds:
\[
\begin{aligned}
&\underset{Y_{0:n,1:m}\sim \check{P}_{0:n,1:m}}{\mathbb{E}}\exp\left(-r_{0,1:m}(\psi)-\sum_{i=1}^n W_ir_{i,1:m}(\psi)\right) \\
&= \int \frac{p(Y_{0,1:m}|\theta_0,\varphi_0)}{\check{p}_{0,1:m}(Y_{0,1:m})}\prod_{i=1}^n\left(\frac{p(Y_{i,1:m}|\tilde{\theta}_i,\varphi_0)}{\check{p}_{i,1:m}(Y_{i,1:m})}\right)^{W_i}\check{P}_{0:n,1:m}(dY_{0:n,1:m}) \\
&= \prod_{i=1}^n \int \left(\frac{p(Y_{i,1:m}|\tilde{\theta}_i,\varphi_0)}{\check{p}_{i,1:m}(Y_{i,1:m})}\right)^{W_i} \check{P}_{i,1:m}(dY_{i,1:m})
\end{aligned}
\]
For an arbitrary single term, it is straightforward to have that
\[
\begin{aligned}
&\int \left(\frac{p(Y_{i,1:m}|\tilde{\theta}_i,\varphi_0)}{\check{p}_{i,1:m}(Y_{i,1:m})}\right)^{W_i} \check{P}_{i,1:m}(dY_{i,1:m}) \\
&=\exp\left(\log\left\lbrace\int (p(Y_{i,1:m}|\tilde{\theta}_i,\varphi_0))^{W_i}(\check{p}_{i,1:m}(Y_{i,1:m}))^{1-W_i} dY_{i,1:m} \right\rbrace \right) \\
&= \exp\left(-m(1-W_i)\mathbb{D}_{W_i}\left(p(\cdot|\tilde{\theta}_i,\varphi_0),\check{p}_i(\cdot)\right) \right).
\end{aligned}
\]
Hence, it follows that
\[
\begin{aligned}
&\underset{Y_{0:n,1:m}\sim \check{P}_{0:n,1:m}}{\mathbb{E}}\exp\left(-r_{0,1:m}(\psi)-\sum_{i=1}^n W_ir_{i,1:m}(\psi)\right) \\
&= \exp\left(-\sum_{i=1}^n m(1-W_i)\mathbb{D}_{W_i}\left(p(\cdot|\tilde{\theta}_i,\varphi_0),\check{p}_i(\cdot)\right) \right).
\end{aligned}
\]
Now move the right hand side term to the left side and multiply $\varepsilon$, so we obtain
\[
\begin{aligned}
&\underset{Y_{0:n,1:m}\sim \check{P}_{0:n,1:m}}{\mathbb{E}}\exp\left\lbrace-r_{0,1:m}(\psi)-\sum_{i=1}^n W_ir_{i,1:m}(\psi)  \right. \\
& \ \ \left. +\sum_{i=1}^n m(1-W_i)\mathbb{D}_{W_i}\left(p(\cdot|\tilde{\theta}_i,\varphi_0),\check{p}_i(\cdot)\right) -\log(\frac{1}{\varepsilon}) \right\rbrace \\
&=\varepsilon.
\end{aligned}
\]
Now we calculate the expectation with respect to prior $\Pi$ and exchange expectations by Fubini's theorem.
\[
\begin{aligned}
&\underset{Y_{0:n,1:m}\sim \check{P}_{0:n,1:m}}{\mathbb{E}} \underset{\psi\sim \Pi}{\mathbb{E}} \exp\left\lbrace-r_{0,1:m}(\psi)-\sum_{i=1}^n W_ir_{i,1:m}(\psi)  \right. \\
& \ \ \left. +\sum_{i=1}^n m(1-W_i)\mathbb{D}_{W_i}\left(p(\cdot|\tilde{\theta}_i,\varphi_0),\check{p}_i(\cdot)\right) -\log(\frac{1}{\varepsilon}) \right\rbrace \\
&=\varepsilon.
\end{aligned}
\]
The Donsker-Varadhan's change of measure states that for any measurable function $\Upsilon:\Psi\rightarrow\mathbb{R}$, we have
\[
\underset{\psi\sim F}{\mathbb{E}} \Upsilon(\psi) \leq \mathbb{D}_{KL}(f(\cdot),\pi(\cdot)) + \log\left( \underset{\psi\sim \Pi}{\mathbb{E}} \exp(\Upsilon(\psi)) \right).
\]
By applying this Donsker-Varadhan's change of measure on the left side of the above equality, we have
\[
\begin{aligned}
&\underset{Y_{0:n,1:m}\sim \check{P}_{0:n,1:m}}{\mathbb{E}} \exp\left\lbrace \underset{\psi\sim F}{\mathbb{E}}\left( -r_{0,1:m}(\psi)-\sum_{i=1}^n W_ir_{i,1:m}(\psi)  \right.\right. \\
& \ \ \left.\left. +\sum_{i=1}^n m(1-W_i)\mathbb{D}_{W_i}\left(p(\cdot|\tilde{\theta}_i,\varphi_0),\check{p}_i(\cdot)\right)\right) -\log(\frac{1}{\varepsilon}) -\mathbb{D}_{KL}(f(\cdot),\pi(\cdot)) \right\rbrace \\
&\leq\varepsilon.
\end{aligned}
\]
By applying the Markov's inequality, with $\check{P}_{0:n,1:m}$ probability at least $(1-\varepsilon)$, we have 
\[
\begin{aligned}
&\exp\left\lbrace \underset{\psi\sim F}{\mathbb{E}}\left( -r_{0,1:m}(\psi)-\sum_{i=1}^n W_ir_{i,1:m}(\psi)  \right.\right. \\
& \ \ \left.\left. +\sum_{i=1}^n m(1-W_i)\mathbb{D}_{W_i}\left(p(\cdot|\tilde{\theta}_i,\varphi_0),\check{p}_i(\cdot)\right)\right) -\log(\frac{1}{\varepsilon}) -\mathbb{D}_{KL}(f(\cdot),\pi(\cdot)) \right\rbrace \\
&\leq 1.
\end{aligned}
\]
Remove the exponential function and multiply $1/m$, we have the following inequality holds
\[
\begin{aligned}
&\int \sum_{i=1}^n (1-W_i) \mathbb{D}_{W_i}\left(p(\cdot|\tilde{\theta}_i,\varphi_0),\check{p}_i(\cdot)\right)F(d\psi) \\
& \ \ \leq \frac{1}{m}\int \left(r_{0,1:m} +\sum_{i=1}^n W_i r_{i,1:m}\right)F(d\psi) + \frac{\mathbb{D}_{KL}(f(\cdot),\pi(\cdot))}{m}+\frac{1}{m}\log\left(\frac{1}{\varepsilon}\right)
\end{aligned}
\]
with $\check{P}_{0:n,1:m}$ probability at least $(1-\varepsilon)$.
\end{proof}

\subsection{Proof of the remark of Theorem 1}
\begin{proof}
Given the inequality in Theorem \ref{THE1}, the left hand side of the inequality can be modified as
\[
\begin{aligned}
& \frac{1}{m}\underset{\psi\sim F}{\mathbb{E}} \left\{ -\sum_{i=1}^n \log\left(\underset{Y_{i,1:m}\sim \check{P}_{i,1:m}}{\mathbb{E}} \left(\frac{p(Y_{i,1:m}|\tilde{\theta}_i,\varphi_0)}{\check{p}_{i,1:m}(Y_{i,1:m})}\right)^{W_i} \right) \right\} \\
&= \underset{\psi\sim F}{\mathbb{E}} \left\{ -\log\left(\underset{Y_{0:n,1:m}\sim \check{P}_{0:n,1:m}}{\mathbb{E}} \exp\left(- m L_{1:m}(\psi)\right)\right)^{\frac{1}{m}} \right\}\\
& \ \ \geq \underset{\psi\sim F}{\mathbb{E}} \left\{ -\log\left(\underset{Y_{0:n,1:m}\sim \check{P}_{0:n,1:m}}{\mathbb{E}} \exp\left(- L_{1:m}(\psi)\right)\right) \right\},
\end{aligned}
\]
and the right hand side of the inequality can be rewritten as:
\[
\underset{\psi\sim F}{\mathbb{E}} L_{1:m}(\psi) +\frac{\mathbb{D}_{KL}(f(\cdot),\pi(\cdot))}{m}+\frac{1}{m}\log\left(\frac{1}{\varepsilon}\right).
\]
Hence, we have derived the ``information posterior bound''.
\end{proof}

\subsection{Proof of the Theorem 2}
\begin{proof}
We rewrite the criterion function as:
\[
\begin{aligned}
M_m(f(\psi)) &= \mathbb{D}_{KL}(f(\cdot),\pi(\cdot)) - \int f(\psi) \log\left(p_{\text{pow}}(Y_{0:n,1:m}|\psi)\right) d\psi.
\end{aligned}
\]
Minimizing $M_m(f(\psi))$ is equivalent to minimizing:
\[
\begin{aligned}
\Delta I(f(\psi)) &= \int f(\psi)\log(f(\psi)) d\psi - \int f(\psi)\log(\pi(\psi)) d\psi + \log(p_{\text{pow}}(Y_{0:n,1:m})) \\
&\ \ -\int f(\psi) \log\left(p_{\text{pow}}(Y_{0:n,1:m}|\psi)\right) d\psi \\
&= \int f(\psi)\log\left(\frac{f(\psi)p_{\text{pow}}(Y_{0:n,1:m})}{p_{\text{pow}}(Y_{0:n,1:m}|\psi) \pi(\psi)}\right) d\psi \\
&= \int f(\psi)\log\left(\frac{f(\psi)}{p_{\text{pow}}(\psi|Y_{0:n,1:m})}\right) d\psi 
\end{aligned}
\]
Obviously we have:
\[
\int p_{\text{pow}}(\psi|Y_{0:n,1:m}) d\psi = \int \frac{p_{\text{pow}}(Y_{0:n,1:m}|\psi) \pi(\psi)}{p_{\text{pow}}(Y_{0:n,1:m})} d\psi =1,
\]
so geographically-powered posterior $P_{\text{pow}}(\psi|Y_{0:n,1:m})$ is a proper probability distribution and therefore we can write $\Delta I(f(\psi))$ as an Kullback-Leibler divergence:
\[
\Delta I(f(\psi)) =  \mathbb{D}_{KL} \left(f(\cdot),\frac{p_{\text{pow}}(Y_{0:n,1:m}|\cdot) \pi(\cdot)}{p_{\text{pow}}(Y_{0:n,1:m})} \right) \geq 0.
\]
It is clear that $f(\psi) = p_{\text{pow}}(\psi|Y_{0:n,1:m})$ minimizes the criterion function $M_m(f(\psi))$ by reducing the difference of input and output information $\Delta I(f(\psi))$ to 0 and thus it results from an optimal information processing rule.
\end{proof}

\subsection{Proof of the Theorem 3}
\begin{proof}
Note that, this theorem easily follows the result of Theorem 1. Here we provide a different way to prove it.

According to Theorem 2, $f(\psi)=p_{\text{pow}}(\psi|Y_{0:n,1:m})$ minimizes $M_m(f(\psi))$. Meanwhile, minimizing $M_m(f(\psi))$ is equivalent to minimizing:
\[
\begin{aligned}
K_m(f(\psi)) &= m^{-1}\int f(\psi)\log\left(\check{p}_{0:n,1:m}(Y_{0:n,1:m})\right)d\psi + m^{-1}\mathbb{D}_{KL}(f(\cdot),\pi(\cdot)) \\
&\ \ -m^{-1}\int f(\psi)\log\left(p(Y_{0,1:m}|\theta_0,\varphi_0)\right)d\psi\\
&\ \ -m^{-1}\sum_{i=1}^n W_i \int f(\psi) \log\left(p(Y_{i,1:m}|\tilde{\theta}_i,\varphi_0)\right)  d\psi \\
&=m^{-1}\int f(\psi) \log\left(\frac{\check{p}_{0:n,1:m}(Y_{0:n,1:m})}{P_{\text{pow}}(Y_{0:n,1:m}|\psi)}\right) d\psi \\
&\ \ + m^{-1}\mathbb{D}_{KL}(f(\cdot),\pi(\cdot)) 
\end{aligned}
\]
Denote the $j$\textsuperscript{th} batch of observations from all locations by $Y_{0:n,j}=(Y_{0,j},Y_{1,j},...,Y_{n,j})$. By independence:
\[
\begin{aligned}
K_m(f(\psi)) &= m^{-1} \int f(\psi) \sum_{j=1}^m \log\left(\frac{\check{p}(Y_{0:n,j})}{p(Y_{0,j}|\theta_0,\varphi_0)\prod_{i=1}^n p(Y_{i,j}|\tilde{\theta}_i,\varphi_0)^{W_i}}\right) d\psi \\
&\ \ + m^{-1}\mathbb{D}_{KL}(f(\cdot),\pi(\cdot)) \\
&= m^{-1} \int f(\psi) \sum_{j=1}^m \left\lbrace \log\left(\frac{\check{p}_0(Y_{i,j})}{p(Y_{i,j}|\theta_0,\varphi_0)}\right)+\sum_{i=1}^n W_i \log\left(\frac{\check{p}_i(Y_{i,j})}{p(Y_{i,j}|\tilde{\theta}_i,\varphi_0)}\right) \right\rbrace d\psi \\
&\ \ + \Lambda^{(m)} + m^{-1}\mathbb{D}_{KL}(f(\cdot),\pi(\cdot)), \\
&= \Lambda^{(m)} + \underset{\psi\sim F}{\mathbb{E}} L_{1:m}(\psi) + m^{-1}\mathbb{D}_{KL}(f(\cdot),\pi(\cdot)),
\end{aligned}
\]
where
\[
\Lambda^{(m)} := m^{-1} \sum_{j=1}^m \left( \sum_{i=1}^n (1-W_i) \log(\check{p}_i(Y_{i,j})\right).
\]
is a constant. Hence we have the geographically-powered posterior $P_{\text{pow}}$ minimizes
\[
\underset{\psi\sim P_{\text{pow}}}{\mathbb{E}} L_{1:m}(\psi) + m^{-1}\mathbb{D}_{KL}(p_{\text{pow}}(\cdot|Y_{0:n,1:m}),\pi(\cdot)).
\]

When $m\rightarrow \infty$, we have $K_m(f(\psi))$ converges to:
\[
\begin{aligned}
K_\infty(f(\psi))&=\int f(\psi)\int \check{p}_{0:n,1:m}(Y_{0:n,1:m})\log\left(\frac{\check{p}_{0:n,1:m}(Y_{0:n,1:m})}{p(y_0|\theta_0,\varphi_0)\prod_{i=1}^n p(y_i|\tilde{\theta}_i,\varphi_0)^{W_i}}\right)dY_{0:n,1:m}d\psi \\
&= \int f(\psi)\int \check{p}_{0:n,1:m}(Y_{0:n,1:m}) \log\left(\frac{\check{p}_0(y_0)}{p(y_0|\theta_0,\varphi_0)}\right)dY_{0:n,1:m}d\psi \\
&\ \ +\int f(\psi)\int \check{p}_{0:n,1:m}(Y_{0:n,1:m}) \sum_{i=1}^n \log\left(\frac{\check{p}_i(y_i)}{p(y_i|\tilde{\theta}_i,\varphi_0)^{W_i}}\right)dY_{0:n,1:m}d\psi.
\end{aligned}
\]
We now look at an arbitrary single term and decompose it:
\[
\begin{aligned}
&\int f(\psi)\int \check{p}_{0:n,1:m}(Y_{0:n,1:m}) \log\left(\frac{\check{p}_i(y_i)}{p(y_i|\tilde{\theta}_i,\varphi_0)^{W_i}}\right)dY_{0:n,1:m}d\psi \\
&= \int f(\psi) (1-W_i) \int \check{p}_i(y_i) \log(\check{p}_i(y_i)) dy_id\psi \\
&\ \ + \int f(\psi) W_i \int \check{p}_i(y_i) \log\left(\frac{\check{p}_i(y_i)}{p(y_i|\tilde{\theta}_i,\varphi_0)}\right)dy_id\psi \\
&= (1-W_i) \int \check{p}_i(y_i) \log(\check{p}_i(y_i)) dy_i + W_i \int f(\psi) \mathbb{D}_{KL}(\check{p}_i(\cdot),p(\cdot|\tilde{\theta}_i,\varphi_0)) d\psi \\
&= (1-W_i) \int \check{p}_i(y_i) \log(\check{p}_i(y_i)) dy_i + W_i \underset{\psi\sim P_{\text{pow}}^{(\infty)}}{\mathbb{E}} \left(\mathbb{D}_{KL}(\check{p}_i(\cdot),p(\cdot|\tilde{\theta}_i,\varphi_0))\right),
\end{aligned}
\]
where the expectation is calculated with respect to distribution $f(\psi)$. We denote a constant $\Lambda$ as:
\[
\Lambda = \sum_{i=1}^n (1-W_i) \int \check{p}_i(y_i) \log(\check{p}_i(y_i)) dy_i.
\]
We then have
\[
\begin{aligned}
K_\infty(f(\psi))&= \Lambda + \underset{\psi\sim P_{\text{pow}}^{(\infty)}}{\mathbb{E}} \left(\mathbb{D}_{KL}(\check{p}_0(\cdot),p(\cdot|\theta_0,\varphi_0))+\sum_{i=1}^n W_i \mathbb{D}_{KL}(\check{p}_i(\cdot),p(\cdot|\tilde{\theta}_i,\varphi_0))\right) \\
&= \Lambda + \underset{\psi\sim P_{\text{pow}}^{(\infty)}}{\mathbb{E}} L(\psi).
\end{aligned}
\]
According to Theorem 2 and assuming the probability measure $P_{\text{pow}}^{(\infty)}$ exists, $f(\psi)=p_{\text{pow}}^{(\infty)}(\psi|Y_{0:n,1:\infty})$ minimizes $K_\infty(f(\psi))$. Since $\Lambda$ is a constant, the geographically-powered posterior $p_{\text{pow}}^{(\infty)}(\psi|Y_{0:n,1:\infty})$ is required to put all its mass at $\psi^\ast=(\theta_0^\ast,\tilde{\theta}_{1:n}^\ast,\varphi_0^\ast)$ when $m\rightarrow\infty$, where $\psi^\ast$ satisfies:
\[
\psi^\ast = \argmin_{\psi=(\theta_0,\tilde{\theta}_{1:n},\varphi_0)} \mathbb{D}_{KL}\left(\check{p}_0(\cdot),p(\cdot|\theta_0,\varphi_0)\right)+\sum_{i=1}^n W_i \mathbb{D}_{KL}\left(\check{p}_i(\cdot),p(\cdot|\tilde{\theta}_i,\varphi_0)\right).
\]
\end{proof}

\subsection{Proof of the Theorem 4}
\begin{proof}
To obtain the best predictive performance for new observations $Y_{0:n}^\ast$ from locations $(u_i,v_i)$, $i=0,...,n$, we need to maximise the expected log pointwise predictive density for $Y_{0:n}^\ast$. Let $\psi_{M_i}$ be the corresponding parameters of model $M_i$, $i=0,...,n$. By the assumption of the geographically weighted regression model (i.e., observation $Y$ is independently generated from the true data generating process, we have
\[
p(Y_{0:n}^\ast|\psi_{M_0},...,\psi_{M_n})=\prod_{i=0}^n p(Y_i^\ast|\psi_{M_0},...,\psi_{M_n}) =\prod_{i=0}^n p(Y_i^\ast|\psi_{M_i}).
\]
By the Assumption 2, we have
\[
p(\psi_{M_0},...,\psi_{M_n}|Y_{0:n,1:m})=\prod_{i=0}^n p_{M_i}(\psi_{M_i}|Y_{0:n,1:m}).
\]
Then we have
\[
\begin{aligned}
p(Y_{0:n}^\ast|Y_{0:n,1:m}) &= \int p(Y_{0:n}^\ast|\psi_{M_0},...,\psi_{M_n}) p(\psi_{M_0},...,\psi_{M_n}|Y_{0:n,1:m}) d\psi_{M_0}...d\psi_{M_n} \\
&= \prod_{i=0}^n \int p(Y_i^\ast|\psi_{M_i}) p_{M_i}(\psi_{M_i}|Y_{0:n,1:m}) d\psi_{M_i}.
\end{aligned}
\]
Plugging it into the expected log pointwise predictive density for $Y_{0:n}^\ast$, we have
\[
\begin{aligned}
\text{elpd}(M) & = \int \log \left(p(Y_{0:n}^\ast|Y_{0:n,1:m})\right) \prod_{i=0}^n \check{p}_i(Y_i^\ast)dY_{0:n}^\ast \\
& = \sum_{i=0}^n \int  \left(\log\left(\int p(Y_i^\ast|\psi_{M_i}) p_{M_i}(\psi_{M_i}|Y_{0:n,1:m}) d\psi_{M_i} \right) \prod_{j=0}^n \check{p}_j(Y_j^\ast) \right)dY_{0:n}^\ast \\
& = \sum_{i=0}^n \int \left( \log\left(\int p(Y_i^\ast|\psi_{M_i}) p_{M_i}(\psi_{M_i}|Y_{0:n,1:m}) d\psi_{M_i} \right)  \check{p}_i(Y_i^\ast) \right)dY_i^\ast \\
& = \sum_{i=0}^n \int \check{p}_i(Y_i^\ast)\log(p_{M_i}(Y_i^\ast|Y_{0:n,1:m}))dY_i^\ast \\
& = \sum_{i=0}^n \text{elpd}_{(u_i,v_i)}(M_i).
\end{aligned}
\]
Given any geographically weighted regression model $M_i=((u_i^\prime,v_i^\prime),\eta)$, by Assumption 1, we have that for $\forall \eta > 0$, $M_i(\eta)=((u_i,v_i),\eta)$ always maximizes the elpd. That is
\[
\text{elpd}_{(u_i,v_i)}(M_i) \leq \text{elpd}_{(u_i,v_i)} (M_i(\eta)),\ i=0,...,n.
\]

Now we have
\[
\text{elpd}(M) \leq \sum_{i=0}^n \text{elpd}_{(u_i,v_i)} (M_i(\eta)).
\]
This has proved $(u_i^\ast,v_i^\ast)=(u_i,v_i)$ for all $i$. To further maximize $\text{elpd}(M)$, we simply require
\[
    \eta^\ast = \argmax_\eta \frac{1}{n+1} \sum_{i=0}^n \text{elpd}_{(u_i,v_i)} (M_i(\eta)) .
\]
\end{proof}

\section{Supplementary Figure}
\begin{figure}[h] 
\setlength{\abovecaptionskip}{0cm}
\setlength{\belowcaptionskip}{0cm}
\includegraphics[width=\textwidth]{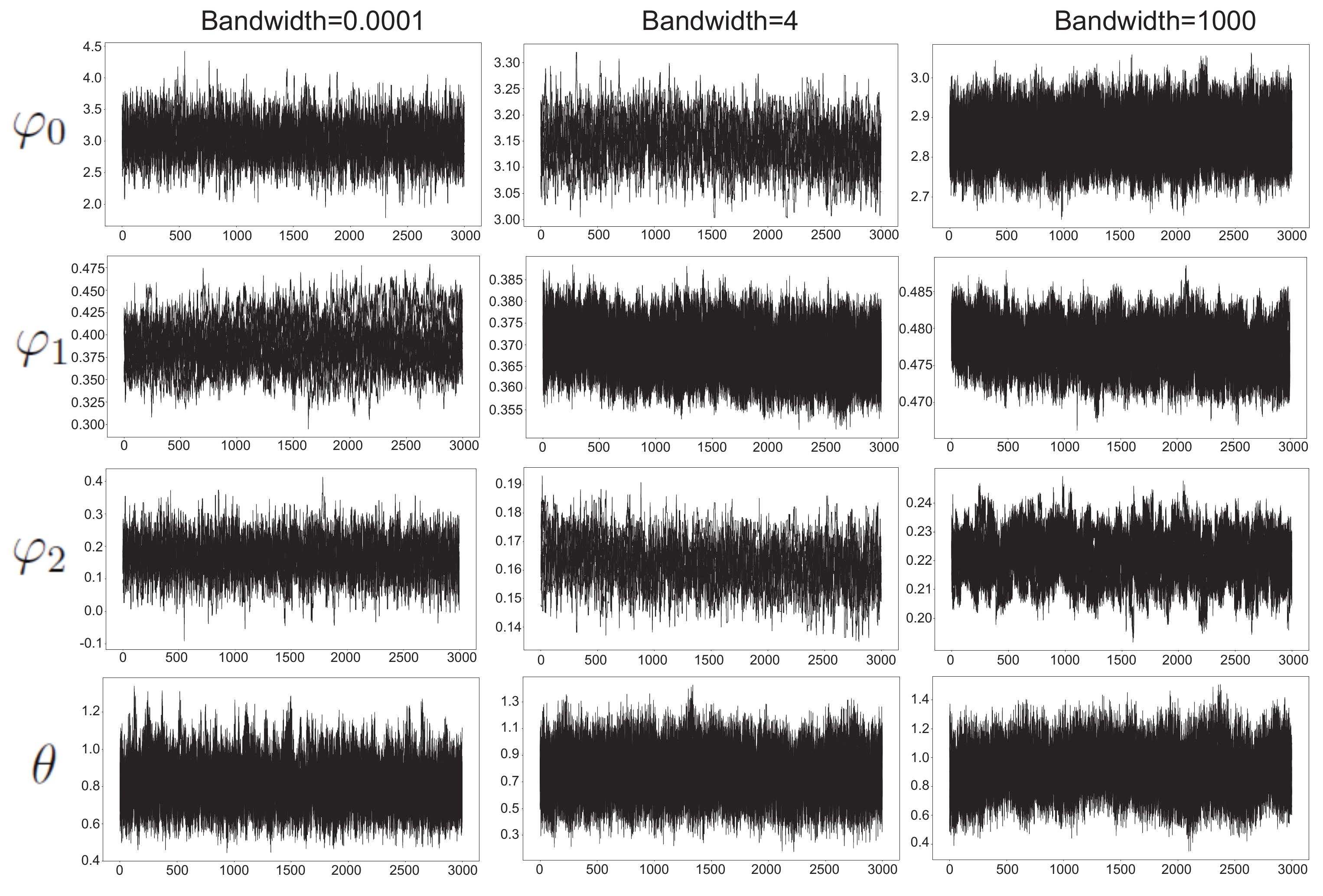} 
\caption[Trace plot of SMI samples for $\varphi_0$, $\varphi_1$, $\varphi_2$ and $\theta$ when $\eta\in\{0.0001,1,20\}$.]{\textbf{Trace plot of SMI samples for $\varphi_0$, $\varphi_1$, $\varphi_2$ and $\theta$ when $\eta\in\{0.0001,1,20\}$.} Each plot contains results of 10 chains. The upper and lower bounds of trace plots reveal that the empirical Bayesian GWR posterior tends to have lower variance with higher geographical bandwidth.} 
\end{figure}

\end{document}